\documentclass[10pt,journal,compsoc]{IEEEtran}
\pdfoutput=1

\usepackage{array}
\usepackage{algorithm,algorithmicx,algpseudocode}
\usepackage{algorithm}
\usepackage{amsfonts} 
\usepackage{amsmath}
\usepackage{amssymb}
\usepackage{array} 
\usepackage[font=small,skip=2pt]{caption}
\ifCLASSOPTIONcompsoc
\usepackage[nocompress]{cite}
\else
\usepackage{cite}
\fi
\usepackage{color} 
\usepackage{enumerate}
\usepackage{floatflt}
\usepackage{float}
\usepackage{footnote}
\usepackage{graphics}
\usepackage[pdftex]{graphicx}
\usepackage{indentfirst}
\usepackage{lineno}
\usepackage{multirow} 
\usepackage[caption=false,font=footnotesize,labelfont=sf,textfont=sf]{subfig}
\usepackage{theorem} 
\usepackage[normalem]{ulem}

\theoremstyle{break}

\newtheorem{theorem}{Theorem}
\newtheorem{lemma}{Lemma}
\newtheorem{corollary}{Corollary}

\newcommand{\ignore}[1]{ }

\newcommand{\beq}{\begin{equation}}
\newcommand{\eeq}{\end{equation}}

\newcommand{\tabincell}[2]{\begin{tabular}{@{}#1@{}}#2\end{tabular}}

\begin{document}
%
\title{An Accurate and Efficient Method to Calculate the Error Statistics of
Block-based\\Approximate Adders}
\author{Yi~Wu$^{\dagger}$, You~Li$^{\dagger}$, Xiangxuan~Ge$^{\dagger}$,
and~Weikang~Qian,~\IEEEmembership{Member,~IEEE}
\IEEEcompsocitemizethanks{
\IEEEcompsocthanksitem Yi Wu, You Li, Xiangxuan Ge, and Weikang Qian are with
the University of Michigan-Shanghai Jiao Tong University Joint Institute,
Shanghai Jiao Tong University, Shanghai, China, 200240.
\IEEEcompsocthanksitem E-mail: \{eejessie, you.li, gxx, qianwk\}@sjtu.edu.cn
\IEEEcompsocthanksitem $^{\dagger}$These authors contributed equally.}}


\IEEEtitleabstractindextext{%
\begin{abstract}
Adders are key building blocks of many error-tolerant applications. Leveraging
the application-level error tolerance, a number of approximate adders were
proposed recently. Many of them belong to the category of block-based
approximate adders. For approximate circuits, besides normal metrics such as
area and delay, another important metric is the error measurement. Given the
popularity of block-based approximate adders, in this work, we propose an
accurate and efficient method to obtain the error statistics of these adders.
We first show how to calculate the error rates. Then, we demonstrate an
approach to get the exact error distribution, which can be used to calculate
other error characteristics, such as mean error distance and mean square error.

\end{abstract}

}

\maketitle

\IEEEdisplaynontitleabstractindextext
\IEEEpeerreviewmaketitle

\ifCLASSOPTIONcompsoc
\IEEEraisesectionheading{\section{Introduction}\label{sec:introduction}}
\else
\section{Introduction}
\label{sec:introduction}
\fi

\IEEEPARstart{A}{pproximate} circuits implement an approximate version of the target function.
They are very attractive for error-tolerant applications, such as image
processing, multimedia, and machine learning, since they can trade off
accuracy for improvement in circuit area, delay, and power
consumption~\cite{han2013approximate}.

Given the importance of adders in building many error-tolerant applications,
approximate adders have attracted a lot of research effort recently. A number
of approximate adders were proposed in literature~\cite{mahdiani2010bio,
gupta2013low, verma2008variable, zhu2009enhanced, kim2013energy,
kahng2012accuracy, hu2015anew, ye2013onreconfig}.  Generally speaking, there
are two design types.  The first type replaces the 1-bit full adders at less
significant bit positions by a simpler but inaccurate module. For example, an
OR gate is used in the Low-Part-OR adder~\cite{mahdiani2010bio} and an
approximate mirror adder is used in~\cite{gupta2013low} to substitute the
accurate 1-bit full adder at the lower bit positions. The more significant part
is intact. As a result, the reduction in delay and power consumption is
limited.  Furthermore, this kind of designs could have a high error rate.

The second design type is known as {\em block-based approximate
adder}~\cite{li2014error}. It divides the entire adder into a number of blocks.
The calculation of the sum in each block exploits the carry speculation
mechanism. It is based on the observation that long carry chain rarely happens
in the addition of random inputs. Therefore, the carry chain for calculating
each sum bit can be truncated at a middle bit position.  Although the carry-in
signal for calculating each sum bit could be wrong, the critical path delay and
the power consumption are reduced. The majority of the available approximate
adders fall into this category.  Examples include the Almost Correct
Adder~\cite{verma2008variable}, the Error Tolerant Adder Type
II~\cite{zhu2009enhanced}, and the Carry-Skip Approximate
Adder~\cite{kim2013energy}. (More details of these approximate adders will be
discussed in Section~\ref{sec:model}). This type of adder generally has a low
error rate. A previous work~\cite{li2014error} also showed that it can achieve
minimum mean error distance under some conditions. Given its popularity and
optimality, we focus on block-based approximate adders in our work.

To measure the performance of an approximate adder, besides the normal metrics
such as area, delay, and power consumption, we also need error statistics,
including error rate, mean error distance, and mean square error. Although the
error statistics of each proposed approximate adder were analyzed by its
authors, the method is ad hoc, depending on the structure of the proposed
adder. Furthermore, not every error metric is given. For example, for some
approximate adders, only error rates were studied, but neither mean error
distance nor mean square error was reported.

To address the above problems, three recent works~\cite{li2014error, 
liu2015analytical,Mazahir16} proposed general methods for 
obtaining important error metrics of block-based approximate adders.
\cite{liu2015analytical} proposed an analytical framework to evaluate the error
statistics of three types of adders: the Almost Correct Adder, the Equal
Segmentation Adder, and the Error Tolerant Adder Type II.  However, its results
are just estimates and different approaches are applied to evaluate different
types of adders.  A more general framework was proposed in~\cite{li2014error},
which can be applied to a wider range of approximate adders. It gives an
accurate analysis on the mean error distance, but the error rate and the mean
square error are still estimates, not exact results. In some cases, the
estimates could be more than 7\% away from the accurate values. Also, none 
of~\cite{li2014error} and~\cite{liu2015analytical} showed how to obtain the
exact error distributions. A recent work~\cite{Mazahir16} provided an accurate
method to obtain the exact error distributions. However, its aim is to provide
an approach to analyze a more general type of approximate adder,
which also includes block-based approximate adder. As a result of making the
approach more general, the method sacrifices efficiency and hence needs a long
runtime in analyzing adders of large sizes. Given that the block-based
approximate adders are one of the most common designs and have good
performance, in this work, we specifically target at this type of adders and
propose an efficient method to obtain their exact error distributions.


Our method works under the assumption that the inputs to the
approximate adder are uniformly distributed. We make this assumption because:
\begin{enumerate}
\item Many approximate adders are not just designed for a specific application.
To estimate the overall performance of an approximate adder over a range of
applications, it is reasonable to assume the inputs are uniformly distributed.
\item For many specific applications, the inputs are more or less close to
uniform distribution.
\item A number of previous works (\cite{mahdiani2010bio, zhu2009enhanced,
kim2013energy, kahng2012accuracy, li2014error, liu2015analytical, du2012high, lin2015high})
in analyzing the error statistics of approximate adders also make this
assumption.
\end{enumerate}

Under the assumption of uniform distribution, we first show an accurate and
efficient method to calculate the error rate. Using this technique, we further
demonstrate an approach to calculate the exact error distribution, by which we
can easily obtain other error metrics of interest, such as mean error distance
and mean square error. Compared to the previous analytical
approaches~\cite{li2014error, liu2015analytical}, our method is able to
generate the exact error distributions and gives exact error characteristics.
Compared to the previous work~\cite{Mazahir16}, our method is much faster
because it exploits the specific properties on the error patterns of the
block-based approximate adders. Indeed, our method achieve the theoretical
lower bound on the asymptotic runtime.
The proposed method provides an important aid to designers in choosing a proper
approximate adder.

In summary, the main contributions of our works are as follows:
\begin{enumerate}
\item We propose an accurate and efficient method to obtain the error rate of
the block-based approximate adders.
\item We propose an efficient method to obtain the exact error distribution of
the block-based approximate adders, which can be used to get other error
characteristics accurately. As we will show, the asymptotic runtime of our
method reaches the theoretical lower bound. 
\item We apply our method to obtain the error distributions of several
previously proposed approximate adders. The results demonstrate the existence
of special patterns on the error distributions of these approximate adders.  We
give explanation for these special patterns.
\item We demonstrate experimentally the proposed method is much faster and more
accurate than the Monte Carlo sampling method to obtain error statistics,
especially when the error probability is very small.
\end{enumerate}

The remainder of the paper is organized as follows. Section~\ref{sec:model}
introduces the general model of the block-based approximate adder and links it
to some previously proposed approximate adders. Section~\ref{sec:preliminary}
discusses some preliminaries. Section~\ref{sec:error-rate} and
Section~\ref{sec:error-distr} show our method to calculate error rate and error
distribution, respectively.  Section~\ref{sec:experiment} presents the
experimental results. Finally, Section~\ref{sec:conclusion} concludes the
paper.

\section{Block-based Approximate Adders}
\label{sec:model}

In this section, we first overview some previously proposed approximate adders.
Then we demonstrate that all of them can be viewed as block-based approximate
adder.

\begin{figure}[!t]
\centering
\includegraphics[scale=0.9]{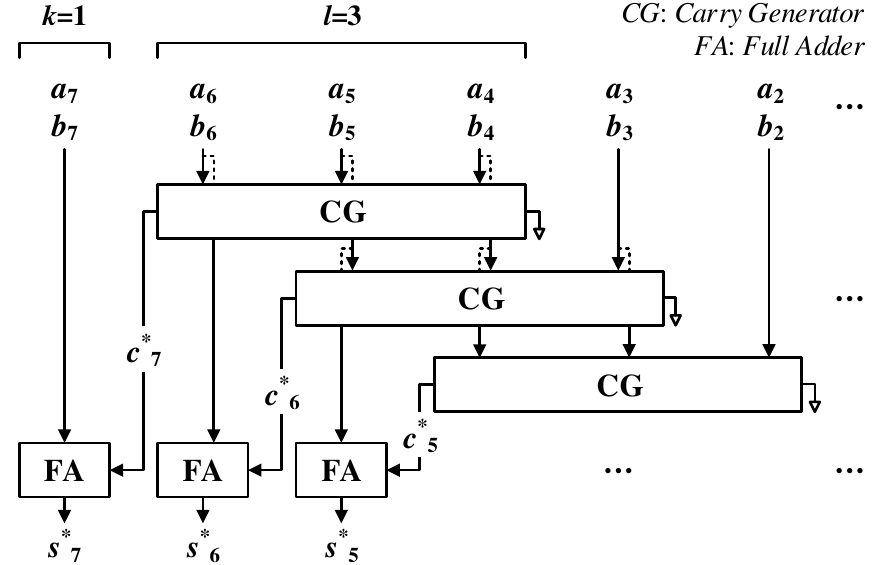}
\caption{\small Almost Correct Adder proposed in~\cite{verma2008variable}.}
\label{fig:aca}
\end{figure}
\begin{figure}[!t]
\centering
\includegraphics[scale=0.9]{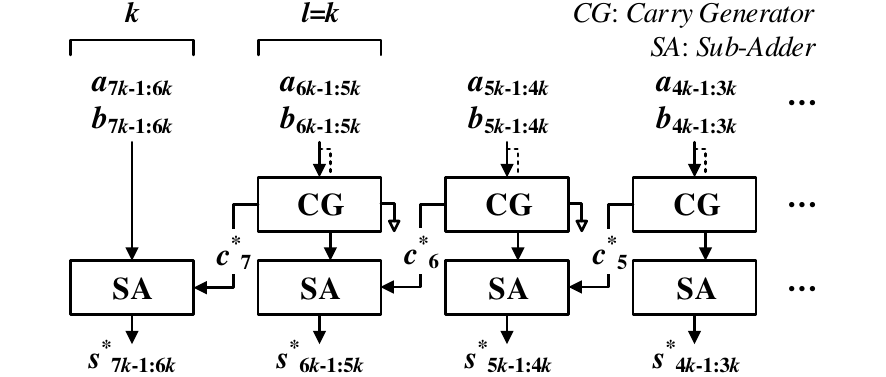}
\caption{\small Error Tolerant Adder Type II proposed
in~\cite{zhu2009enhanced}.}
\label{fig:eta2}
\end{figure}

The adder proposed in~\cite{lu2004speeding} and the Almost Correct Adder
(ACA)~\cite{verma2008variable} has a structure shown in Fig.~\ref{fig:aca}.
Each sum bit is produced by a full adder, which takes a carry-in as input.
However, the carry-in is only obtained from $l$ bits before the sum bit instead
of all the remaining bits. The Error Tolerant Adder Type II (ETA-II), as shown
in Fig.~\ref{fig:eta2}, divides the entire $n$-bit adder into $m$ sub-adders of
equal bit length of $k=n/m$~\cite{zhu2009enhanced}. The carry-in signal to each
sub-adder is produced from the previous $k$ bits by a carry generator, while
the carry-in to each carry generator is a logic $0$, which essentially
truncates the carry chain. Other block-based approximate adders include
Speculative Carry Select Adder (SCSA)~\cite{du2012high}, Error Tolerant Adder
Type IV (ETA-IV)~\cite{zhu2010enhanced}, and Carry-Skip Approximate Adder
(CSAA)~\cite{kim2013energy}. At the behavior level, given the same sub-adder
length, SCSA is the same as ETA-II. ETA-IV is similar to ETA-II except that the
length of its sub-adder is twice that of its carry generator.  CSAA is also
similar to ETA-II except that the length of its carry generator is twice that
of its sub-adder.

\begin{figure}[!t]
\centering
\includegraphics[scale=0.9]{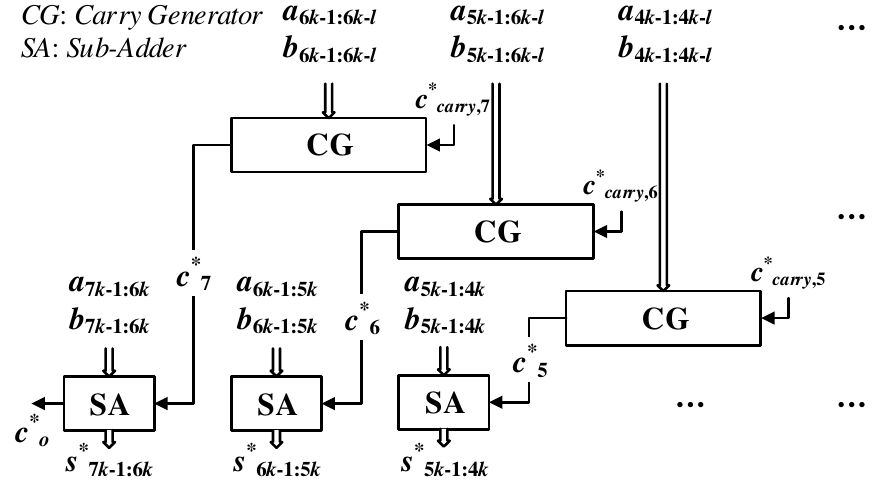}
\caption{\small General model of a block-based approximate adder.}
\label{fig:general-model}
\end{figure}

All of the above-mentioned approximate adders can be viewed as block-based
approximate adder~\cite{li2014error}, with a general model shown in
Fig.~\ref{fig:general-model}. Notice that a similar model was also proposed
in~\cite{shafique2015alow}. In the model, the number of bits is $n$. Assume the
two addends of an $n$-bit adder are $A=a_{n-1} \ldots a_0$ and $B=b_{n-1}
\ldots b_0$. The approximate sum is denoted as $S^*=s_{n-1}^*, \ldots s_0^*$.
The carry-out of the approximate adder is denoted as $c_o^*$. In the model, the
sum is divided into a number of blocks which are calculated separately. All the
blocks are with the same bit length of $k$, where $k$ is a factor of $n$.  Let
$m = n/k$, which represents the number of blocks. In the model, the sum bits in
the $i$-th ($0 \le i \le m-1$) block, $s_{(i+1)k-1}^* \ldots s_{ik}^*$, are
generated by a sub-adder, which takes a speculated carry-in $c_{i}^*$. In the
ideal case, the carry-in should be produced by all the input bits lower than
the position $ik$.  However, for the block-based approximate adder, $c_{i}^*$
is produced by a truncated carry generator of length $l$, as shown in
Fig.~\ref{fig:general-model}. The carry generator also takes a speculated
carry-in $c_{carry,i}^*$.  For most of the approximate adders, $c_{carry,i}^* =
0$.  Thus, in the following analysis, we will assume that $c_{carry,i}^* = 0$,
although our analysis is equally applicable to the case where $c_{carry,i}^* =
1$. Note that for all $0 \le i \le \lfloor l/k \rfloor$, the speculated
carry-in $c_i^*$ is produced by all the remaining input bits and hence, it is
always correct. The carry-out $c_o^*$ of the entire adder is produced by the
leftmost sub-adder.

In summary, a block-based approximate adder is characterized by 3 parameters,
$n$, $k$, and $l$, where $n$ is the adder size, $k$ is the block size, and $l$
is the number of bits used in the carry generator.  All of the above-mentioned
approximate adders are just special cases of this model. For example, the adder
proposed in~\cite{lu2004speeding} and ACA correspond to the case where $k=1$.
ETA II and SCSA correspond to the case where $k=l$.  ETA-IV corresponds to the
case where $k=2l$. CSAA corresponds to the case where $l=2k$.

Generally speaking, block-based approximate adder can be extended to one with
different sub-adder lengths and carry generator lengths for different blocks,
which is the one considered in~\cite{Mazahir16}. However, many existing
approximate adders have the same sub-adder length and the same carry-generator
length for all the blocks, since for this kind of design, no particular block
dominates the critical path length. Given this fact, in this work, we target at
the block-based approximate adders. For these adders, as we will show in
Section~\ref{subsec:time_complexity}, our method to obtain the exact error
distribution has the lowest asymptotic runtime. However, it should be pointed
out that our proposed method can be easily extended to handle the more general
situation, at the cost of increasing the asymptotic runtime.

\section{Preliminaries}
\label{sec:preliminary}

In this section, we show some preliminaries that will be used in our later
analysis. We assume the inputs $A$ and $B$ are uniformly distributed in $[0,
2^n-1]$ and the carry-in to the entire adder is $0$.

\subsection{Propagate, Generate, and Kill Signals}

For each bit $i$ ($0 \le i \le n-1$) in the adder, the propagate, generate, and
kill signals of that bit are defined as
\begin{equation*}
p_i = a_i \oplus b_i, g_i = a_i \cdot b_i, k_i = \bar{a_i} \cdot \bar{b_i}.
\end{equation*}

If $g_i = 1$, the carry-out of bit $i$ is 1 regardless what the carry-in to bit
$i$ is. Similarly, if $k_i=1$, the carry-out of bit $i$ is always 0. If $p_i =
1$, then the carry-in of bit $i$ propagates to the carry-out of that bit.

For the $i$-th ($0 \le i \le m-1$) block of the adder, we define the group
propagate, generate, and kill signal as
\begin{eqnarray*}
& P_i = \prod_{j=ik}^{(i+1)k-1} p_j, \\
& G_i = \sum_{j=ik}^{(i+1)k-1} g_j \prod_{d=j+1}^{(i+1)k-1} p_d, \\
& K_i = \sum_{j=ik}^{(i+1)k-1} k_j \prod_{d=j+1}^{(i+1)k-1} p_d.
\end{eqnarray*}

If $G_i = 1$, the carry-out of the $i$-th block will always be the
correct value of 1 no matter its carry-in is correct or not. Similarly, if $K_i
= 1$, the carry-out will always be the correct value of 0. Only when $P_i = 1$
does the carry-out depend on the carry-in signal, which could be wrong.  The
probabilities of the above signals being one are
\begin{eqnarray}
\label{eqn:p-p-i}
& P(P_i) \stackrel{\triangle}{=} P(P_i =1) = \frac{1}{2^k}, \\
\label{eqn:p-g-i}
& P(G_i) \stackrel{\triangle}{=} P(G_i =1) = \frac{1}{2} - \frac{1}{2^{k+1}},
\\
\label{eqn:p-k-i}
& P(K_i) \stackrel{\triangle}{=} P(K_i =1) = \frac{1}{2} - \frac{1}{2^{k+1}}.
\end{eqnarray}

\subsection{Typical Error Measurement}

Typical error measurement of an approximate arithmetic circuit includes error
rate, mean error distance, and mean square error. 

First, we define the error distance (ED) as the difference of the approximate
sum $S^*$ and the accurate sum $S$, i.e.,
$$ED=\left|S^*-S\right|.$$

The error rate (ER) is defined as the percentage of input combinations for
which the approximate adder produces a wrong result, i.e., a non-zero error
distance. Mathematically, it is calculated as
\begin{equation*}
ER=P(ED\neq0).
\end{equation*}

Mean error distance (MED) is the mean value of all the error distances.
Mean square error (MSE) are the mean value of the squares of all
the error distances. Mathematically, they are calculated as
\begin{eqnarray*}
MED=E[ED]=\sum_{ED_i \in \Omega} ED_{i}P(ED_i),\\
MSE=E[ED^2]=\sum_{ED_i \in \Omega} ED_{i}^2 P(ED_i),
\end{eqnarray*}
where $\Omega$ is the set of all error distances.

\section{Calculating Error Rate}
\label{sec:error-rate}

In this section, we show the method to calculate the error rate. It will be
used later to obtain the exact error distribution.

As can be seen in Fig.~\ref{fig:general-model}, the result of the approximate
adder is correct if and only if all the speculated carry-in $c_i^*$'s ($0 \le i
\le m-1$) are correct. To calculate the error rate, we define the event $D_i$
as the event in which all the speculated carry-ins $c_i^*, c_{i-1}^*, \ldots,
c_0^*$ are correct.  We denote the probability of the event $D_i$ to occur as
$d_i$. Thus, the error rate equals $1-d_{m-1}$.  In the following, we will
derive a recursive formula to calculate $d_i$.
We denote the correct carry-in to the $i$-th sub-adder as $c_i$. Since the
recursive formula differs based on whether or not the carry generator length
$l$ is a multiple of the block size $k$, we will distinguish these two cases.


\subsection{Carry Generator Length $l$ is a Multiple of Block Size $k$}
\label{sec:ER-k-equal}

\begin{figure}[!t]
\centering
\includegraphics[scale=0.85]{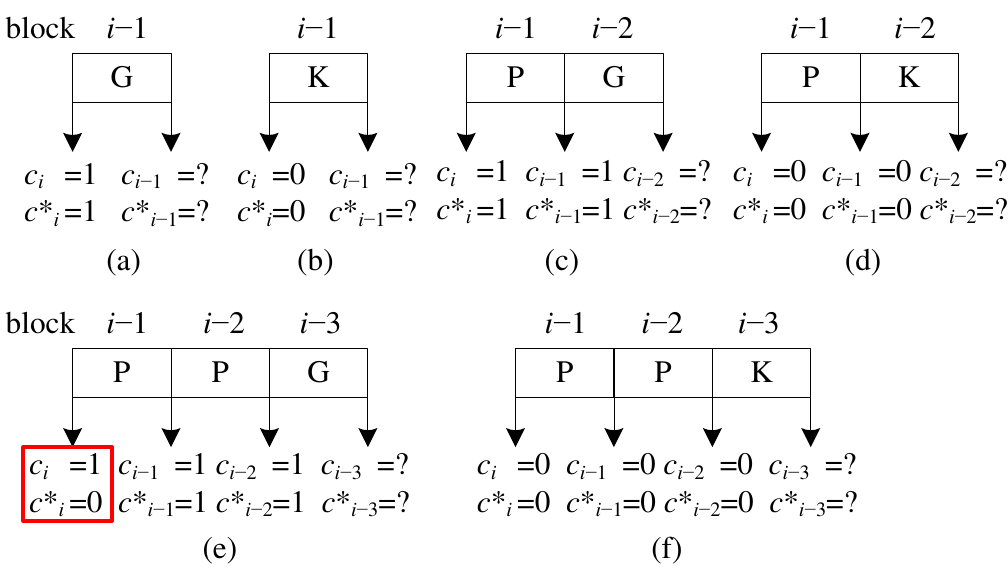}
\caption{\small The speculated carry-ins and the correct carry-ins for
different input cases under the situation that $l = 2k$. (a) $G_{i-1} = 1$;
(b) $K_{i-1}=1$; (c) $P_{i-1} = G_{i-2} = 1$; (d) $P_{i-1} = K_{i-2} = 1$;
(e) $P_{i-1} = P_{i-2} = G_{i-3} = 1$; (f)  $P_{i-1} = P_{i-2} = K_{i-3} = 1$.}
\label{fig:error-rate}
\end{figure}

Define $t = \frac{l}{k}$. Then $t$ represents the number of blocks covered by
each carry generator.  To illustrate our proposed method, we use $t = 2$ as an
example.  In this case, the carry generator includes 2 blocks of inputs. For $0
\le i \le 2$, since all the remaining input bits are used to generate the
speculated carry-ins $c_i^*, c_{i-1}^*, \ldots, c_0^*$, the event $D_i$ always
happens.  Thus, $d_i = 1$, for all $0 \le i \le 2$.

Next we consider $d_i$ for $i >2$. The event $D_i$ depends on the inputs from
block $i-1$ to $0$. Our idea to calculate $d_i$ is to consider the inputs block
by block from block $i-1$ to block $0$.

First consider the inputs at block $i-1$. They satisfy either $G_{i-1} = 1$,
$K_{i-1} = 1$, or $P_{i-1} = 1$. If the inputs satisfy that $G_{i-1} = 1$, as
shown in Fig.~\ref{fig:error-rate}(a), the speculated carry-in $c_i^* = 1$,
which is equal to the correct carry-in $c_{i}$. Thus, the event $D_i$ happens
if and only if $c_{i-1}^*, \ldots, c_0^*$ are correct, which means the inputs
from block $i-2$ to $0$ make the event $D_{i-1}$ happen. Therefore,
we have
\begin{equation}
\label{eqn:mul-g}
P(D_i, G_{i-1}=1)= P(G_{i-1}) P(D_{i-1}).
\end{equation}

The same conclusion holds if the inputs at block $i-1$ satisfy that $K_{i-1} =
1$, as shown in Fig.~\ref{fig:error-rate}(b). Therefore, we have
\begin{equation}
\label{eqn:mul-k}
P(D_i, K_{i-1}=1)= P(K_{i-1}) P(D_{i-1}).
\end{equation}

If the inputs at block $i-1$ satisfy $P_{i-1} = 1$, then we further consider
the inputs at block $i-2$. We also distinguish them into three cases:
$G_{i-2} = 1$, $K_{i-2} = 1$, and $P_{i-2} = 1$. In the case where
$G_{i-2}=1$ (shown in Fig.~\ref{fig:error-rate}(c)) and the case where
$K_{i-2}=1$ (shown in Fig.~\ref{fig:error-rate}(d)), since the carry generator
covers 2 blocks of inputs, the speculated carry-ins $c_i^*$ and $c_{i-1}^*$ are
equal to the correct carry-ins $c_i$ and $c_{i-1}$, respectively.  Thus, the
event $D_i$ happens if and only if $c_{i-2}^*, \ldots, c_0^*$ are correct,
which means the inputs from block $i-3$ to $0$ make the event $D_{i-2}$
happen. Therefore, we have
\begin{eqnarray}
\label{eqn:mul-p-g}
P(D_i, P_{i-1}=G_{i-2}=1) = P(P_{i-1}) P(G_{i-2}) P(D_{i-2}), \\
\label{eqn:mul-p-k}
P(D_i, P_{i-1}=K_{i-2}=1) = P(P_{i-1}) P(K_{i-2}) P(D_{i-2}).
\end{eqnarray}

If the inputs at blocks $i-1$ and $i-2$ satisfy none of the above cases, then
we must have $P_{i-1} = P_{i-2} = 1$. Now we further consider the inputs at
block $i-3$. We distinguish the following three cases:
\begin{enumerate}
\item The inputs satisfy that $G_{i-3}=1$, as shown in
Fig.~\ref{fig:error-rate}(e). In this case, the correct carry-ins $c_i =
c_{i-1} = c_{i-2} = 1$. However, the speculated carry-in $c_i^*$ is 0, since it
is produced by a carry generator that covers inputs at blocks $i-1$ and $i-2$
and that carry generator propagates the speculated carry-in to the carry
generator, which is assumed to be 0. Since $c_i^* \neq c_i$, the event $D_i$
cannot happen in this case. Therefore, we have
\begin{equation}
\label{eqn:mul-p2-g}
P(D_i, P_{i-1}= P_{i-2}= G_{i-3}=1) = 0.
\end{equation}
\item The inputs satisfy that $K_{i-3}=1$, as shown in
Fig.~\ref{fig:error-rate}(f). In this case, the correct carry-ins $c_i =
c_{i-1} = c_{i-2} = 0$. By the same argument used in Case 1, the
speculated carry-in $c_i^*$ must be 0. Since each carry generator covers two
blocks of inputs, the speculated carry-in $c_{i-1}^* = c_{i-2}^* = 0$.
Therefore, $c_j^* = c_j$ for $j = i, i-1, i-2$. Thus, the event $D_i$ happens
if and only if $c_{i-3}^*, \ldots, c_0^*$ are correct, which means the inputs
from block $i-4$ to $0$ make the event $D_{i-3}$ happen. Therefore, we have
\begin{equation}
\label{eqn:mul-p2-k}
\begin{split}
&P(D_i, P_{i-1}= P_{i-2}= K_{i-3}=1) \\
& = P(P_{i-1}) P(P_{i-2}) P(K_{i-3}) P(D_{i-3}).
\end{split}
\end{equation}
\item The inputs satisfy that $P_{i-3}=1$. In this case, we continue to look
at the inputs at block $i-4$.
\end{enumerate}

We continue the above analysis. By the same reasoning used for the case where
$P_{i-1} = P_{i-2} = G_{i-3} = 1$, we have that for any $3 < j \le i$,  if the
inputs from block $i-1$ to block $i-j$ satisfy that $P_{i-1} = \cdots =
P_{i-j+1} = G_{i-j} = 1$, the event $D_i$ cannot happen, since $c_i^*=0 \neq
c_i = 1$, i.e.,
\begin{equation}
\label{eqn:mul-pn-g}
P(D_i, P_{i-1}= P_{i-2}= \cdots = P_{i-j+1} = G_{i-j}=1) = 0.
\end{equation}

On the other hand, if
the inputs satisfy that $P_{i-1} = \cdots = P_{i-j+1} = K_{i-j} = 1$, the
event $D_i$ happens if and only if the inputs from block $i-j-1$ to $0$ make
the event $D_{i-j}$ happen. Therefore, we have
\begin{equation}
\label{eqn:mul-kn-g}
\begin{split}
&P(D_i, P_{i-1}= P_{i-2}= \cdots = P_{i-j+1} = K_{i-j}=1) \\
& = P(P_{i-1}) P(P_{i-2}) \cdots P(P_{i-j+1}) P(K_{i-j}) P(D_{i-j}).
\end{split}
\end{equation}

Finally, there is a remaining input case which
satisfies that $P_{i-1} = \cdots = P_0 = 1$. In this case, the speculated
carry-ins are $c_{i-1}^* = \cdots = c_0^* = 0$ and the correct carry-ins are
$c_{i-1} = \cdots = c_0 = 0$.  Thus, the event $D_i$ happens. Therefore,
we have
\begin{equation}
\label{eqn:mul-all-p}
\begin{split}
&P(D_i, P_{i-1}= P_{i-2}= \cdots = P_{0}=1)\\
&= P(P_{i-1}) P(P_{i-2}) \cdots P(P_{0}).
\end{split}
\end{equation}

Notice the probability that the event $D_i$ occurs can be calculated as
\begin{small}
\begin{equation*}
\begin{split}
P(D_i) &= P(D_i, G_{i-1}=1) + P(D_i, K_{i-1}=1) \\
&+ P(D_i, P_{i-1} = G_{i-2}=1) + P(D_i, P_{i-1} = K_{i-2} = 1)\\
&+ \cdots + P(D_i, P_{i-1} = \cdots = P_1 = G_0=1) \\
&+ P(D_i, P_{i-1} = \cdots = P_1 = K_0=1) \\
&+ P(D_i, P_{i-1} = \cdots = P_1 = P_0=1).
\end{split}
\end{equation*}
\end{small}

Given Eq.~\eqref{eqn:mul-g}--\eqref{eqn:mul-all-p}, we can calculate $d_i$ as
follows:
\begin{small}
\begin{equation*}
\begin{split}
d_i &= P(D_i)
= P(G_{i-1}) P(D_{i-1}) + P(K_{i-1}) P(D_{i-1}) \\
&+ P(P_{i-1}) P(G_{i-2}) P(D_{i-2}) + P(P_{i-1}) P(K_{i-2}) P(D_{i-2}) \\
&+ P(P_{i-1}) P(P_{i-2}) P(K_{i-3}) P(D_{i-3}) \\
&+ P(P_{i-1}) P(P_{i-2}) P(P_{i-3}) P(K_{i-4}) P(D_{i-4}) \\
&+ \cdots + P(P_{i-1}) P(P_{i-2}) \cdots P(P_1) P(K_0) P(D_{0}) \\
&+ P(P_{i-1}) P(P_{i-2}) \cdots P(P_0) \\
= &P(P_{i-1}) \cdots P(P_0) + \sum_{j=1}^2 P(P_{i-1}) \cdots P(P_{i-j+1})
P(G_{i-j}) d_{i-j} \\
&+ \sum_{j=1}^i P(P_{i-1}) \cdots P(P_{i-j+1}) P(K_{i-j}) d_{i-j}.
\end{split}
\end{equation*}
\end{small}

For an arbitrary $t$, we can generalize the above analysis and obtain that for
$0 \le i \le t$, $d_i = 1$ and for $i > t$,
\begin{equation*}
\begin{split}
d_i &= P(P_{i-1}) \cdots P(P_0) \\
&+ \sum_{j=1}^t P(P_{i-1}) \cdots P(P_{i-j+1})
P(G_{i-j}) d_{i-j} \\
&+ \sum_{j=1}^i P(P_{i-1}) \cdots P(P_{i-j+1}) P(K_{i-j}) d_{i-j}.
\end{split}
\end{equation*}

The above equation gives a recursive way to calculate $d_i$. The values of
$P(P_i), P(G_i), P(K_i)$ are calculated by Eq.~\eqref{eqn:p-p-i},
\eqref{eqn:p-g-i}, and~\eqref{eqn:p-k-i}, respectively. The time complexity to
obtain $d_{m-1}$ and the resultant error rate is $O(m^2)$.

\subsection{Carry Generator Length $l$ is not a Multiple of Block Size $k$}
\label{sec:ER-k-neq}

Define $t = \lfloor \frac{l}{k} \rfloor$. Each carry speculative chain is
composed of $t$ blocks of bit length $k$ and a remaining block of bit length
$k' = l-tk$. Note that $0 < k' < k$.

For each block of $k$ bits, we divide it into the left group of $k'$ bits and
the right group of $k-k'$ bits. The major difference in the analysis here is
that we need to consider the propagate/generate/kill state of both the left
group and the right group in a block. For the left group of $k'$ bits in the
$i$-th ($0 \le i \le m-1$) block of the adder, we define its group propagate,
generate, and kill signal as
\begin{eqnarray*}
& PL_i = \prod_{j=(i+1)k-k'}^{(i+1)k-1} p_j, \\
& GL_i = \sum_{j=(i+1)k-k'}^{(i+1)k-1} g_j \prod_{d=j+1}^{(i+1)k-1} p_d, \\
& KL_i = \sum_{j=(i+1)k-k'}^{(i+1)k-1} k_j \prod_{d=j+1}^{(i+1)k-1} p_d.
\end{eqnarray*}

The probabilities of the above signals being one are
\begin{eqnarray*}
& P(PL_i) \stackrel{\triangle}{=} P(PL_i =1) = \frac{1}{2^{k'}}, \\
& P(GL_i) \stackrel{\triangle}{=} P(GL_i =1) = \frac{1}{2} -
\frac{1}{2^{k'+1}}, \\
& P(KL_i) \stackrel{\triangle}{=} P(KL_i =1) = \frac{1}{2} -
\frac{1}{2^{k'+1}}.
\end{eqnarray*}

Similarly, we define the group propagate, generate, and kill signal of the
right group of $k-k'$ bits of the $i$-th block. These signals are denoted as
$PR_i$, $GR_i$, and $KR_i$, respectively. Their probabilities of being one are
calculated similarly as above.

\begin{figure}[!t]
\centering
\includegraphics[scale=0.85]{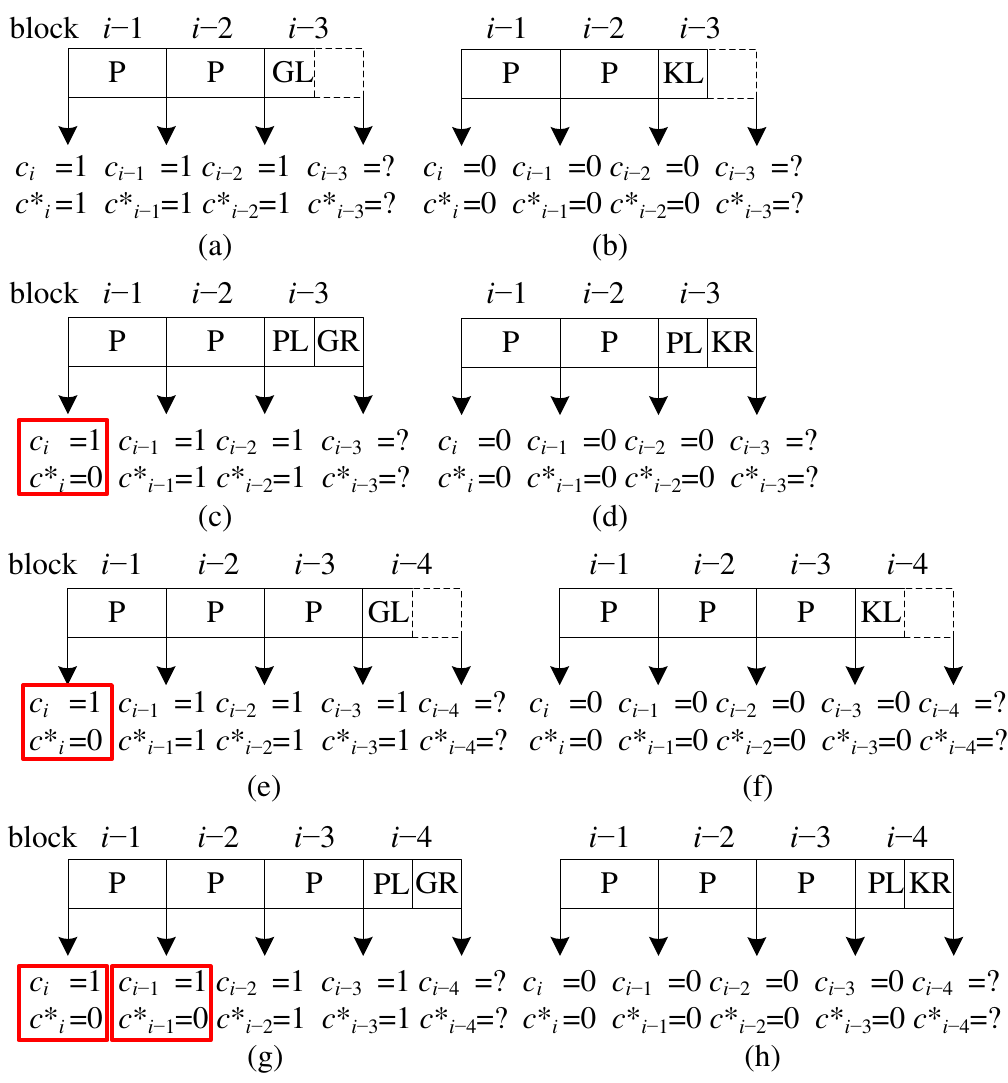}
\caption{\small The speculated carry-ins and the correct carry-ins for
different input cases under the situation that $l = 2k+k'$ where $0 < k' < k$.
(a) $P_{i-1} = P_{i-2} = GL_{i-3} = 1$; (b) $P_{i-1} = P_{i-2} = KL_{i-3} = 1$;
(c) $P_{i-1} = P_{i-2} = PL_{i-3} = GR_{i-3}= 1$; (d) $P_{i-1} = P_{i-2} =
PL_{i-3} = KR_{i-3}= 1$.}
\label{fig:error-rate-general}
\end{figure}

To illustrate the method to calculate $d_i$, we also use $t = 2$ as an example.
For $0 \le i \le 2$, it is not hard to see that $d_i = 1$. Thus, we only focus
on $i > 2$.

The basic idea to obtain the probability $d_i$ is same as what we use to handle
the case where $l$ is a multiple of $k$, i.e., examining inputs block by block
from block $i-1$ to block $0$.  For the inputs satisfying either (1) $G_{i-1} =
1$, (2) $K_{i-1} = 1$, (3) $P_{i-1} = G_{i-2} = 1$, or (4) $P_{i-1} = K_{i-2} =
1$, the conclusions are the same as what we have when $l = 2k$.
Fig.~\ref{fig:error-rate}(a)-(d) show the correct carry-ins and the speculated
carry-ins for these four cases, respectively.

If the inputs at blocks $i-1$ and $i-2$ satisfy none of the above cases,
then we have $P_{i-1} = P_{i-2} = 1$. We further consider the inputs at block
$i-3$. The difference compared to the situation where $l$ is a multiple of $k$
is that we need to distinguish the following five cases:
\begin{enumerate}
\item The inputs satisfy that $GL_{i-3} = 1$, as shown in
Fig.~\ref{fig:error-rate-general}(a). Since each carry generator covers two
blocks of inputs plus the left group of the third block, the speculated
carry-ins $c_i^* = c_{i-1}^* = c_{i-2}^* = 1$. In the correct adder, the
carry-ins $c_i = c_{i-1} = c_{i-2} = 1$. Thus, the event $D_i$ happens if and
only if $c_{i-3}^*, \ldots, c_0^*$ are correct, which means the inputs from
block $i-4$ to $0$ make the event $D_{i-3}$ happen. Therefore, we have
\begin{equation}
\label{eqn:not-mul-p2-gl}
\begin{split}
&P(D_i, P_{i-1}= P_{i-2}= GL_{i-3}=1) \\
& = P(P_{i-1}) P(P_{i-2}) P(GL_{i-3}) P(D_{i-3}).
\end{split}
\end{equation}
\item The inputs satisfy that $KL_{i-3} = 1$, as shown in
Fig.~\ref{fig:error-rate-general}(b). The analysis is same as Case 1 and we
have the same conclusion: the event $D_i$ happens if and only if the inputs
from block $i-4$ to $0$ make the event $D_{i-3}$ happen. Therefore, we have
\begin{equation}
\label{eqn:not-mul-p2-kl}
\begin{split}
&P(D_i, P_{i-1}= P_{i-2}= KL_{i-3}=1) \\
& = P(P_{i-1}) P(P_{i-2}) P(KL_{i-3}) P(D_{i-3}).
\end{split}
\end{equation}
\item The inputs satisfy that $PL_{i-3} = GR_{i-3} = 1$, as shown in
Fig.~\ref{fig:error-rate-general}(c). In this case, the correct carry-ins $c_i
= c_{i-1} = c_{i-2} = 1$. However, the speculated carry-in $c_i^*$ is 0, since
it is produced by a carry generator that covers inputs at block $i-1$, block
$i-2$, and the left group of block $i-3$ (see
Fig.~\ref{fig:error-rate-general}(c) and that carry generator propagates a 0.
Since $c_i^* \neq c_i$, the event $D_i$ cannot happen. Therefore, we have
\begin{equation}
\label{eqn:not-mul-p2-pl-gr}
\begin{split}
&P(D_i, P_{i-1}= P_{i-2}= PL_{i-3} = GR_{i-3} = 1) = 0.
\end{split}
\end{equation}
\item The inputs satisfy that $PL_{i-3} = KR_{i-3} = 1$. In this case, as
shown in Fig.~\ref{fig:error-rate-general}(d), $c_j^* = c_j$ for $j = i-1,
i-2, i-3$. Thus, the event $D_i$ happens if and only if the inputs from
block $i-4$ to $0$ make the event $D_{i-3}$ happen. Therefore, we have
\begin{equation}
\label{eqn:not-mul-p2-pl-kr}
\begin{split}
&P(D_i, P_{i-1}= P_{i-2}= PL_{i-3}= KR_{i-3}=1) \\
& = P(P_{i-1}) P(P_{i-2}) P(PL_{i-3}) P(KR_{i-3}) P(D_{i-3}).
\end{split}
\end{equation}
\item The inputs satisfy that $P_{i-3} = 1$. In this case, we need to continue
checking the inputs at block $i-4$.
\end{enumerate}

Now we consider the remaining case where $P_{i-1}=P_{i-2}=P_{i-3} = 1$. We
further check the inputs at block $i-4$. Similarly, they can be divided into
the five cases as shown above. The situations corresponding to the first four
cases are shown in Fig.~\ref{fig:error-rate-general}(e)-(h), respectively.
Since $P_{i-1} = P_{i-2}=P_{i-3} = 1$ and each carry generator covers two
blocks of inputs plus the left group of the third block, the speculated
carry-in $c_i^* = 0$. In Case 1 (i.e., $GL_{i-4}=1$) and Case 3 (i.e.,
$PL_{i-4} = GR_{i-4} = 1$), since the correct carry-in $c_i =1 \neq c_i^*$, the
event $D_i$ cannot happen. Therefore we have
\begin{small}
\begin{eqnarray}
\label{eqn:not-mul-p3-gl}
P(D_i, P_{i-1}= P_{i-2}= P_{i-3}= GL_{i-4}=1) = 0, \\
\label{eqn:not-mul-p3-pl-gr}
P(D_i, P_{i-1}= P_{i-2}= P_{i-3}= PL_{i-4} = GR_{i-4}=1) = 0.
\end{eqnarray}
\end{small}
In Case 2 (i.e., $KL_{i-4}=1$) and Case 4 (i.e.,
$PL_{i-4} = KR_{i-4} = 1$), $c_j^* = c_j$ for $j = i-1, \ldots, i-4$. Thus, the
event $D_i$ happens if and only if the inputs from block $i-5$ to $0$ make the
event $D_{i-4}$ happen. Therefore we have
\begin{small}
\begin{eqnarray}
\label{eqn:not-mul-p3-kl}
\begin{split}
& P(D_i, P_{i-1}= P_{i-2}= P_{i-3}= KL_{i-4}=1) \\
& = P(P_{i-1}) P(P_{i-2}) P(P_{i-3}) P(KL_{i-4}) P(D_{i-4}).
\end{split} \\
\label{eqn:not-mul-p3-pl-kr}
\begin{split}
&P(D_i, P_{i-1}= P_{i-2}= P_{i-3}= PL_{i-4} = KR_{i-4}=1) \\
&= P(P_{i-1}) P(P_{i-2}) P(P_{i-3}) P(PL_{i-4}) P(KR_{i-4}) P(D_{i-4}).
\end{split}
\end{eqnarray}
\end{small}
In Case 5, the inputs at block $i-4$ satisfy that
$P_{i-4}=1$; we continue analyzing the inputs of the next block in the same
way.

By the same reasoning used for the case where $P_{i-1} = P_{i-2} = P_{i-3} =
1$, we have that for any $4 < j \le i$, if the inputs from block $i-1$ to block
$i-j$ satisfy either $P_{i-1} = \cdots = P_{i-j+1} = GL_{i-j} = 1$  or
$P_{i-1} = \cdots = P_{i-j+1} = PL_{i-j} = GR_{i-j} = 1$, the event $D_i$
cannot happen. If the inputs satisfy either $P_{i-1} = \cdots = P_{i-j+1} =
KL_{i-j} = 1$ or $P_{i-1} = \cdots = P_{i-j+1} = PL_{i-j} = KR_{i-j} = 1$, the
event $D_i$ happens if and only if the inputs from block $i-j-1$ to $0$ make
the event $D_{i-j}$ happen. The equations to calculate the probabilities are
similar to Eq.~\eqref{eqn:not-mul-p3-kl} and \eqref{eqn:not-mul-p3-pl-kr}.
Finally, for the remaining input case in which $P_{i-1} = \cdots = P_0 = 1$,
the event $D_i$ happens.

By the above discussion, for the example in which $t=2$, we can calculate
$d_i$ as follows:
\begin{equation*}
\begin{split}
d_i &= P(D_i) = P(G_{i-1}) P(D_{i-1}) + P(K_{i-1}) P(D_{i-1}) \\
&+ P(P_{i-1}) P(G_{i-2}) P(D_{i-2}) + P(P_{i-1}) P(K_{i-2}) P(D_{i-2}) \\
&+ P(P_{i-1}) P(P_{i-2}) P(GL_{i-3}) P(D_{i-3}) \\
&+ \sum_{j=3}^{i} [P(P_{i-1}) \cdots P(P_{i-j+1}) P(KL_{i-j}) P(D_{i-j}) \\
&+ P(P_{i-1}) \cdots P(P_{i-j+1}) P(PL_{i-j}) P(KR_{i-j}) P(D_{i-j})] \\
&+ P(P_{i-1}) P(P_{i-2}) \cdots P(P_0) \\
\end{split}
\end{equation*}

Noticing that $P(KL_{j}) + P(PL_{j}) P(KR_{j}) = P(K_j)$, we can further
simplify the above equation as
\begin{equation*}
\begin{split}
d_i &= \sum_{j=1}^2 P(P_{i-1}) \cdots P(P_{i-j+1}) P(G_{i-j}) d_{i-j} \\
&+ \sum_{j=1}^i P(P_{i-1}) \cdots P(P_{i-j+1}) P(K_{i-j}) d_{i-j} \\
&+ P(P_{i-1}) P(P_{i-2}) P(GL_{i-3}) d_{i-3}\\
&+ P(P_{i-1}) \cdots P(P_0).
\end{split}
\end{equation*}

For an arbitrary $t$, we can generalize the above analysis and obtain that for
$0 \le i \le t$, $d_i = 1$ and for $i > t$,
\begin{equation*}
\begin{split}
d_i &= \sum_{j=1}^t P(P_{i-1}) \cdots P(P_{i-j+1}) P(G_{i-j}) d_{i-j} \\
&+ \sum_{j=1}^i P(P_{i-1}) \cdots P(P_{i-j+1}) P(K_{i-j}) d_{i-j} \\ &+
P(P_{i-1}) \cdots P(P_{i-t}) P(GL_{i-t-1}) d_{i-t-1} \\
&+ P(P_{i-1}) \cdots
P(P_0).
\end{split}
\end{equation*}

The above equation gives a recursive way to calculate $d_i$. The time
complexity to obtain $d_{m-1}$ and the resultant error rate is $O(m^2)$.

\section{Obtaining Error Distribution}
\label{sec:error-distr}

In this section, we show our method to obtain the accurate error distribution.
For a given input combination, the {\em output error} is defined as $err =
c_o^* s_{n-1}^* \ldots s_0^*- c_o s_{n-1} \ldots s_0$, where $c_o, s_{n-1},
\ldots, s_0$ are the correct outputs of the adder. By the
definition, the error distance of an error is equal to the absolute value of
the error. The error distribution is the probability distribution of the error
distance. Similar as characterizing the error rate, we will discuss based on
whether or not the carry generator length $l$ is a multiple of the block size
$k$ in Section~\ref{subsec:mul} and~\ref{subsec:not_mul}, respectively.
Finally, we will discuss the time complexity of our method in
Section~\ref{subsec:time_complexity}.

\subsection{Carry Generator Length $l$ is a Multiple of Block Size $k$}\label{subsec:mul}

\subsubsection{Error Pattern and Probability}

We define $t = l/k$.  In this section, we first analyze the {\em error
pattern}, which is the binary representation of the error distance.  The
questions we ask are: 1) when does an error happen? 2) if an error happens,
what does its error pattern look like?

\begin{figure}[!htbp]
	\centering
	\includegraphics[scale=0.9]{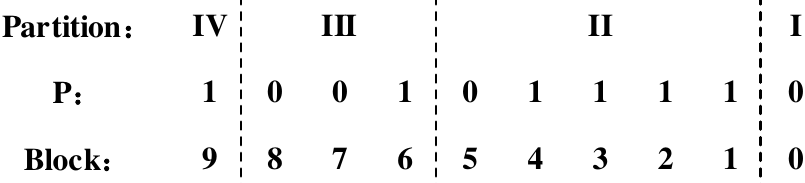}
	\caption{\small An example of 10 blocks divided into 4 partitions. }
	\label{fig:patition}
\end{figure}

To analyze the error pattern, we divide the $m$ blocks into a number of
partitions. Each partition is composed of two sequences of blocks, where all
the blocks in the left sequence have their group propagate signals as 0 and all
the blocks in the right sequence have their group propagate signals as 1. For
the leftmost (rightmost) partition, it is possible that all of its blocks have
their group propagate signals as 1 (0). Fig.~\ref{fig:patition} gives an
example of 10 blocks divided into 4 partitions.

We now study the approximate sum in each partition.
Suppose the partition starts at block $i_b$ and ends at block $i_e$ ($i_b >
i_e$). The blocks $i_b, i_b-1, \ldots, i_m$ ($i_e < i_m \le i_b$) have their
group propagate signals as $0$ and the blocks $i_m-1, \ldots, i_e$ have their
group propagate signals as $1$.  We have the following two claims on errors in
the partition.

\begin{lemma}
\label{lem:error-partition-cond}
There exists an error in the approximate sum of a partition if and only if
$G_{i_e-1} = 1$ and $i_m - i_e \ge t$.
\end{lemma}

\begin{IEEEproof}
``if'' part: The carry speculative chain for the carry-in $c_{i_m}^*$ is
composed of the bits in blocks $i_m-1, \ldots, i_m-t$. Since $P_{i_m-1} =
\cdots = P_{i_m-t} = 1$, we have $c_{i_m}^* = 0$. However, because $P_{i_m-1} =
\cdots = P_{i_e} = 1$ and $G_{i_e-1} = 1$ , the correct carry-in $c_{i_m} = 1$.
Thus, the sum at block $i_m$ is incorrect.

``only if'' part: we prove by contradiction. If the condition is not satisfied,
then there are two cases: (1) $K_{i_e-1} = 1$ and (2) $G_{i_e-1} = 1$ and $i_m
- i_e < t$. For Case 1, the speculated carry-ins $c_{i_m}^*, \ldots, c_{i_e}^*$
are all equal to the correct value of 0. For Case 2, the speculated carry-ins
$c_{i_m}^*, \ldots, c_{i_e}^*$ are all equal to the correct value of 1.  For
both cases, since $P_{i_b-1} = \cdots = P_{i_m} = 0$, the speculated carry-ins
$c_{i_b}^*, \ldots, c_{i_{m}+1}^*$ are all correct. Therefore, for both cases,
the speculated carry-ins $c_{i_b}^*, \ldots, c_{i_e}^*$ are all correct and
hence, the approximate sum of the partition is correct.
\end{IEEEproof}

\begin{lemma}
\label{lem:error-partition-pattern}
If there exists an error in the approximate sum of a partition, the approximate
sum from block $(i_e+t-1)$ to block $i_e$, i.e., $s_{(i_e+t)k-1}^* \ldots
s_{i_e k}^*$, is correct and the approximate sum from block $i_b$ to block
$(i_e+t)$, i.e., $s_{(i_b+1)k-1}^* \ldots s_{(i_e+t)k}^*$, is smaller than the
correct sum by $1$.
\end{lemma}

\begin{IEEEproof}
If there exists an error in the approximate sum of a partition, by
Lemma~\ref{lem:error-partition-cond}, $P_{i_e+t-1} = \cdots = P_{i_e} = 1$, and
$G_{i_e-1} = 1$.  Since each carry speculative chain is composed of $t$ blocks,
the speculated carry-ins $c_{i_e+t-1}^*, \ldots, c_{i_e}^*$ equal the correct
value of $1$. Therefore, the approximate sum from block $(i_e+t-1)$ to block
$i_e$ is correct.

Since $P_{i_m-1} = \cdots = P_{i_e+t-1} = 1$, the correct carry-ins $c_{i_m} =
\ldots = c_{i_e+t} = 1$. However, due to the truncation of the carry
speculative chain, the speculated carry-ins $c_{i_m}^* = \ldots = c_{i_e+t}^* =
0$. Furthermore, considering the fact that $P_{i_m-1} = \cdots = P_{i_e+t} =
1$, the correct sums of blocks $i_m-1, \ldots, i_e+t$ are $k$ 0's in the binary
representation, while the approximate sums of these blocks are $k$ 1's in the
binary representation. Since $P_{i_m} = 0$, $c_{i_m} = 1$, and $c_{i_m}^* = 0$,
the approximate sum of block $i_m$ is smaller than its correct sum by 1.
Finally, since $P_{i_b-1} = \cdots = P_{i_m} = 0$, the speculated carry-ins
$c_{i_b}^*, \ldots, c_{{i_m}+1}^*$ are all correct and hence the sums of blocks
$i_b, \ldots, i_m+1$ are all correct. Given the above characterization of the
sums of blocks $i_b, \ldots, i_e+t$, we can see that the approximate sum from
block $i_b$ to block $(i_e+t)$ is smaller than the correct sum by $1$.
\end{IEEEproof}



\begin{figure}[!t]
\centering
\includegraphics[scale=0.95]{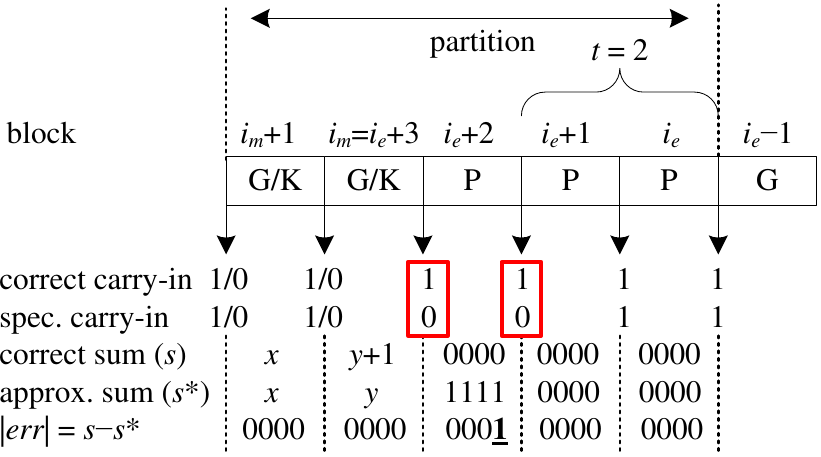}
\caption{\small Illustration of the error pattern $|err|$ in a partition. }
\label{fig:error-in-partition}
\end{figure}

An illustration of
Lemma~\ref{lem:error-partition-pattern} is shown in
Fig.~\ref{fig:error-in-partition}. In this example, we assume the block size is
$k=4$, the number of blocks covered by each carry generator is $t = 2$, $i_m =
i_e+3$, and $i_b = i_m+1$. Then, the correct carry-ins $c_{i_e+3} = c_{i_e+2} =
c_{i_e+1} = c_{i_e} = 1$, while the speculated carry-ins $c_{i_e+3}^* =
c_{i_e+2}^* = 0$ and $c_{i_e+1}^* = c_{i_e}^* = 1$.  For block $i_m$, since its
group propagate signal $P_{i_m} \neq 1$, we have $c_{i_m+1}^* = c_{i_m+1}$.
Therefore, the approximate sums at blocks $i_m+1$, $i_e+1$, and $i_e$ are
correct. The approximate sum at block $i_m$ is smaller than the correct sum by
1.  The approximate sum at block $i_e+2$ is $(1111)_2$, while the correct sum
at that block is $(0000)_2$. Overall, the approximate sum from block $i_m+1$ to
block $i_e+2$ is smaller than the correct value by 1, while the approximate sum
from block $i_e+1$ to block $i_e$ is correct.

By Lemma~\ref{lem:error-partition-pattern}, the error of an approximate adder
is always negative and the $1$'s in an error pattern can only occur at
the bit position $ik$, where $i$ is an integer.  Furthermore,
Lemma~\ref{lem:error-partition-pattern} implies that if a $1$ appears at a
position $ik$, then the block $i$ must be inside a partition whose rightmost
block is block $i-t$. By Lemma~\ref{lem:error-partition-cond}, we should have
$P_{i-1} = \cdots = P_{i-t} = 1$ and $G_{i-t-1} = 1$. Therefore, we have the
following claim.

\begin{theorem}
\label{thm:err-cond}
A 1 in an error pattern can only occur at the bit position $ik$, where $i$ is
an integer. It occurs at the bit position $ik$ if and only if $P_{i-1} = \cdots
= P_{i-t} = 1$ and $G_{i-t-1} = 1$.
\end{theorem}

%

Based on the above theorem, we also have the following corollary.
\begin{corollary}
\label{cor:err-exclusive}
If there are two adjacent 1's in an error pattern, where the left one is at
position $ik$ and the right one is at position $jk$, then $j < i-t$.
\end{corollary}

\begin{IEEEproof}
Since a 1 occurs at the bit position $ik$ of the error pattern, by
Theorem~\ref{thm:err-cond}, we have $G_{i-t-1}=1$. As a result, 1 cannot occur
at bit positions $(i-1)k, \ldots, (i-t)k$, because otherwise, due to
Theorem~\ref{thm:err-cond}, we have $P_{i-t-1}=1$, which contradicts with
$G_{i-t-1}=1$.  Therefore, the next 1 in the error pattern must be at the bit
position $jk$, where $j < i-t$.
\end{IEEEproof}

\begin{figure}[!t]
\centering
\includegraphics[scale=0.90]{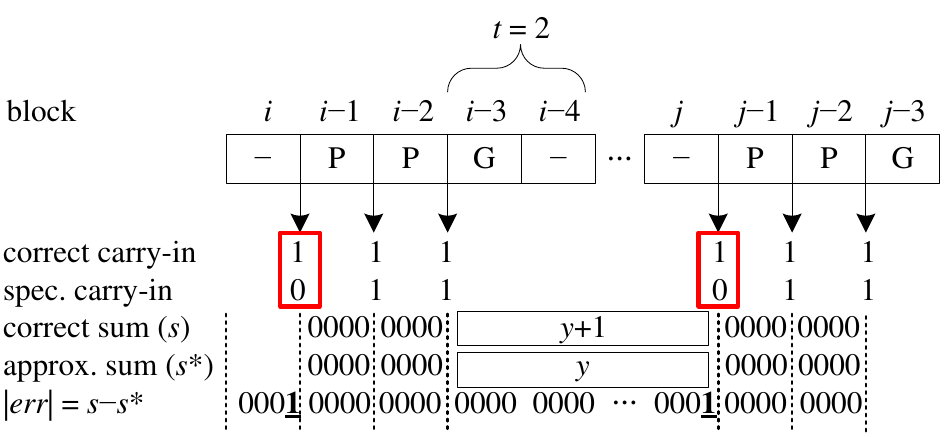}
\caption{\small Illustration of the situation in which a $1$ appears at bit
position $jk$ of the error pattern $|err|$ and the next 1 appears at bit
position $ik$ ($i > j$). Here we assume $k=4$ and $t=2$.}
\label{fig:adjacent-error}
\end{figure}

Now we study the following question: if there is a 1 at position $jk$ of an
error pattern, under what situation does the next 1 on the left of $jk$ appear
at a position $ik$ where $i > j+t$? Fig.~\ref{fig:adjacent-error} illustrates
this problem using an example with $k=4$ and $t=2$. As shown in the figure, in
the error pattern, there is a 1 at bit position $jk$ and the next $1$ is at bit
position $ik$.  Since the correct carry-in to the bit $jk$ is 1, the correct
sum from block $i-3$ to block $j$ is
\begin{equation}
\label{eqn:correct-sum}
s_{(i-2)k-1} \ldots s_{jk} = a_{(i-2)k-1} \ldots a_{jk} + b_{(i-2)k-1} \ldots
b_{jk}+1,
\end{equation}
where $a_{(i-2)k-1} \ldots a_{jk}$ and $b_{(i-2)k-1} \ldots b_{jk}$ denote
the inputs from block $i-3$ to block $j$.

On the other hand, as shown in Fig.~\ref{fig:adjacent-error}, except the last
bit of block $j$, all the other bits in blocks $i-3, \ldots, j$ of the error
pattern $|err|$ are $0$. Therefore, the correct sum from block $i-3$ to $j$ is
larger than the approximate sum by 1.
Given Eq.~\eqref{eqn:correct-sum}, the approximate sum is
\begin{equation}
\label{eqn:approx-sum}
s_{(i-2)k-1}^* \ldots s_{jk}^* = a_{(i-2)k-1} \ldots a_{jk} + b_{(i-2)k-1}
\ldots b_{jk}.
\end{equation}


Furthermore, the approximate sum from block $i-3$ to $j$ is equal to the sum
produced by an approximate adder of the same type with $(i-j-2)$ blocks, where
the inputs are $a_{(i-2)k-1} \ldots a_{jk}$ and $b_{(i-2)k-1} \ldots b_{jk}$.
For clarity, we call this approximate adder {\em imaginary
approximate adder}.  Given Eq.~\eqref{eqn:approx-sum}, we have that the
inputs $a_{(i-2)k-1} \ldots a_{jk}$ and $b_{(i-2)k-1} \ldots b_{jk}$ should
make the imaginary approximate adder produce the correct sum. This means all
the speculated carry-ins $c_{i-j-3}^*, \ldots, c_0^*$ of the imaginary adder
should be correct. In other words, the inputs at blocks $i-4, \ldots, j$ of the
adder in Fig.~\ref{fig:adjacent-error} should make the Event $D_{i-j-3}$
defined in Section~\ref{sec:error-rate} happen. For a general $t$, the inputs
at blocks $i-t-2, \ldots, j$ should make the Event $D_{i-t-j-1}$ happen.


Therefore, given that there is a 1 at position $jk$ of an error pattern, the
next 1 appears at a position $ik$ where $i > j+t$ if and only if  $P_{i-t} =
\cdots = P_{i-1} = 1$, $G_{i-t-1} = 1$, and the inputs at blocks $i-t-2,
\ldots, j$ make the event $D_{i-t-j-1}$ happen. The probability is
\begin{equation}
\label{eqn:e-i-j}
e_{i,j} \stackrel{\triangle}{=}  P(P_{i-1}) \cdots P(P_{i-t}) P(G_{i-t-1})
d_{i-t-j-1}.
\end{equation}

The rightmost position in the error pattern where a $1$ can occur is the
position $(t+1)k$.  The probability that the rightmost $1$ is at a bit position
$ik$ $(i \ge t+1)$ is
\begin{equation*}
e_{i,0}  \stackrel{\triangle}{=} P(P_{i-1}) \cdots P(P_{i-t}) P(G_{i-t-1})
d_{i-t-1}.
\end{equation*}

The leftmost position in the error pattern where a $1$ can occur is the
position $(m-1)k$. Given that there is a 1 at position $ik$ $(i \le m-1)$,
there is no 1 on the left of position $ik$ if and only if the inputs at blocks
$m-2, \ldots, i$ make the event $D_{m-i-1}$ happen. The probability is
$d_{m-i-1}$.

Given the above discussion, we finally get the following claim on error
pattern and probability.

\begin{theorem}
\label{thm:error-distribution}
A $mk$-bit binary representation is a possible error pattern if and only if
there exist numbers $t+1 \le i_1 < i_2 < \cdots < i_r \le m-1$ such that for
all $1 \le j \le r-1$, we have $i_{j+1} - i_j >t$ and the ones in the binary
representation appear at bit positions $i_1 k, \ldots, i_r k$. The probability
of that error pattern is
\begin{equation*}
d_{m-i_r-1} \cdot \prod_{j=1}^r e_{i_j ,i_{j-1}},
\end{equation*}
where $i_0 = 0$.
\end{theorem}

\subsubsection{Algorithm to Obtain Error Distribution}
\label{sec:impl}

Using Theorem~\ref{thm:error-distribution}, we can enumerate all possible error
patterns and calculate their probabilities. This gives us a method to obtain
the error distribution.

However, the enumeration-based method can be further optimized by saving common
multiplications of probabilities and common additions of error components
appeared in the calculation. We propose a divide-and-conquer method to do this.
The idea is to grow a partial error pattern and its probability into the
complete error pattern and the probability. The procedure uses a recursive
helper function $ED(i,j, ePar, pPar)$ shown in Algorithm~\ref{alg:error-distr}.
The argument $i$ refers to the bit position $ik$, which is under check and can
potentially have a 1. $j$ refers to the bit position $jk$, which is the nearest
bit position on the right of $ik$ in the error pattern that has a 1.  $ePar$ is
the partial error distance and $pPar$ is the partial probability.

When checking the block $i$, we have the following two cases:
\begin{enumerate}
\item There is a $1$ at the bit position $ik$ in the error pattern. Then, the
nearest bit position on the left of $ik$ that could have a $1$ is $(i+t+1)k$.
We continue to check that bit position. The error magnitude $2^{ik}$ is added
to the partial error distance
and the partial probability is multiplied with the value
$e_{i,j}$. This is shown in Line~\ref{line:i-i+t+1} of
Algorithm~\ref{alg:error-distr}.
\item There is no $1$ at the bit position $ik$ in the error pattern. Then, we
further check the bit position $(i+1)k$. The partial error distance and
probability do not change. This is shown in Line~\ref{line:i-j} of
Algorithm~\ref{alg:error-distr}.
\end{enumerate}

\begin{algorithm}[!htbp]
\begin{small}
\caption{\small $ED(i, j, ePar, pPar)$: a recursive helper function to obtain
the error distribution.
}
\label{alg:error-distr}
\begin{algorithmic}[1]
\If{$i \ge m$}
	\State $pPar = pPar \cdot d_{m-j-1}$;
	\State Print out $ePar$ and $pPar$;
	\State \Return;
\EndIf
\State $ED(i+t+1, i, ePar+2^{ik}, pPar \cdot e_{i,j})$;
\label{line:i-i+t+1}
\State $ED(i+1, j, ePar, pPar)$;
\label{line:i-j}
\State \Return;
\end{algorithmic}
\end{small}
\end{algorithm}

When $i \ge m$, $ePar$ becomes the complete error distance. The probability of
$ePar$ should be $pPar$ multiplied by $d_{m-j-1}$, which accounts for the
probability that there is no 1 on the left of position $jk$ in the error
pattern.  The initial function call is \mbox{$ED(t+1, 0, 0, 1)$}, because the
rightmost position in an error pattern where a $1$ can occur is the position
$(t+1)k$, the partial error distance is 0, and the partial probability is 1.

\subsection{Carry Generator Length $l$ is not a Multiple of Block Size $k$}\label{subsec:not_mul}

This situation is slightly more complicated. However, the overall analysis flow
is similar to the case where $l$ is a multiple of $k$.
Similar as Section~\ref{sec:ER-k-neq}, we define $t = \lfloor \frac{l}{k}
\rfloor$ and $k' = l - tk$. In this situation,
Lemma~\ref{lem:error-partition-cond} is changed to the following one.

\begin{lemma}
\label{lem:error-partition-cond-k-neq}
There exists an error in the approximate sum of a partition if and only if
either of the following two events happens:
\begin{enumerate}
\item
$PL_{i_e-1} = GR_{i_e-1} = 1$ and $i_m - i_e \ge t$, where $PL_i$ and $GR_i$
are defined in Section~\ref{sec:ER-k-neq}.
\item
$GL_{i_e-1} = 1$ and $i_m - i_e \ge t+1$.
\end{enumerate}
\end{lemma}

\begin{IEEEproof}
``if'' part: If the event 1 happens, then the speculated carry-in $c_{i_e+t}^*
= 0$, because the carry generator for the carry-in $c_{i_e+t}^*$ covers blocks
$i_e+t-1, \ldots, i_e$ plus the left group of block $i_e-1$, and their group
propagate signals are all 1. On the other hand, the correct carry-in $c_{i_e+t}
= 1$. Thus, the sum at block $(i_e+t)$ is incorrect. By a similar argument, if
the event 2 happens, then the speculated carry-in $c_{i_e+t+1}^* = 0$, while
the correct carry-in $c_{i_e+t+1} = 1$. Thus, the sum at block $(i_e+t+1)$ is
incorrect.

``only if'' part: we prove by contradiction. If neither of the two events
happens, then there are four cases: (1) $KL_{i_e-1}=1$, (2) $PL_{i_e-1} =
KR_{i_e-1} = 1$, (3) $PL_{i_e-1} = GR_{i_e-1}=1$ and $i_m - i_e \le t-1$, and
(4) $GL_{i_e-1} = 1$ and $i_m - i_e \le t$.  For all cases, the speculated
carry-ins $c_{i_b}^*, \ldots, c_{i_e}^*$ are all correct and hence, the
approximate sum of the partition is correct.
\end{IEEEproof}
As a result, Theorem~\ref{thm:err-cond} is changed to the following one.

\begin{theorem}
\label{thm:err-cond-k-neq}
A 1 in an error pattern can only occur at the bit position $ik$, where $i$ is
an integer. It occurs at the bit position $ik$ if and only if either of the
following two events happens:
\begin{enumerate}
\item
$P_{i-1} = \cdots = P_{i-t} = 1$, $PL_{i-t-1} = 1$, and $GR_{i-t-1} = 1$.
\item
$P_{i-1} = \cdots = P_{i-t-1} = 1$ and $GL_{i-t-2} = 1$.
\end{enumerate}
\end{theorem}


Given the above theorem, we can see that if two adjacent 1's in an error
pattern are at positions $ik$ and $jk$ ($i>j$), respectively, then $i-j > t$.
For the case where $l$ is not a multiple of $k$, we still use $e_{i,j}$ to
denote the probability that given there is a 1 at position $jk$ in an error
pattern, the next 1 appears at a position $ik$, where $i > j+t$. However, to
obtain $e_{i,j}$, we need to distinguish between $i=j+t+1$ and $i>j+t+1$.

Given that there is a 1 at position $jk$ in an error pattern, the next 1
appears at a position $ik$ where $i = j+t+1$ if and only if $P_{i-1} = \cdots =
P_{i-t} = 1$, $PL_{i-t-1} = 1$, and $GR_{i-t-1} = 1$. The probability is
\begin{equation*}
e_{i,i-t-1} \stackrel{\triangle}{=}  P(P_{i-1}) \cdots P(P_{i-t}) P(PL_{i-t-1})
P(GR_{i-t-1}).
\end{equation*}

Given that there is a 1 at position $jk$ in an error pattern, the next 1
appears at a position $ik$ where $i > j+t+1$ if and only if either of the
following two events happens:
\begin{enumerate}
\item
$P_{i-1} = \cdots = P_{i-t} = 1$, $PL_{i-t-1} = 1$, $GR_{i-t-1} = 1$, and the
inputs at blocks $i-t-2, \ldots, j$ make the event $D_{i-t-j-1}$ happen.
\item
$P_{i-1} = \cdots = P_{i-t-1} = 1$, $GL_{i-t-2} = 1$, and the inputs at blocks
$i-t-3, \ldots, j$ make the event $D_{i-t-j-2}$ happen.
\end{enumerate}
The probability that one of the above two events happens is
\begin{equation*}
\begin{split}
e_{i,j} &\stackrel{\triangle}{=} P(P_{i-1}) \cdots P(P_{i-t}) P(PL_{i-t-1})
P(GR_{i-t-1}) d_{i-t-j-1} \\
&+ P(P_{i-1}) \cdots P(P_{i-t-1}) P(GL_{i-t-2}) d_{i-t-j-2}.
\end{split}
\end{equation*}

The rightmost position in an error pattern where a $1$ can occur is the
position $(t+1)k$.  It can be shown that the probability that the rightmost $1$
is at the position $(t+1)k$ is
\begin{equation*}
e_{t+1,0} \stackrel{\triangle}{=}  P(P_{t}) \cdots P(P_1) P(PL_0) P(GR_0),
\end{equation*}
while the probability that the rightmost $1$ is at the position
$ik$ $(i > t+1)$ is
\begin{equation*}
\begin{split}
e_{i,0} &\stackrel{\triangle}{=} P(P_{i-1}) \cdots P(P_{i-t}) P(PL_{i-t-1})
P(GR_{i-t-1}) d_{i-t-1} \\
&+ P(P_{i-1}) \cdots P(P_{i-t-1}) P(GL_{i-t-2}) d_{i-t-2}.
\end{split}
\end{equation*}


The probability that given there is a 1 at position $ik$ $(i \le m-1)$ in the
error pattern, there is no 1 on the left of position $ik$ is still $d_{m-i-1}$,
the same value as we have for the case where $l$ is a multiple of $k$.

For the case where $l$ is not a multiple of $k$,
Theorem~\ref{thm:error-distribution} still holds, which characterizes the error
pattern and probability. The only difference is that the values $e_{i,j}$ are
replaced with the new  $e_{i,j}$ obtained in this section.  Finally, the
procedure to obtain the error distribution is similar to that used for the case
where $l$ is a multiple of $k$. Again, the only change is to use the new values
of $e_{i,j}$ obtained in this section.

\subsection{Time Complexity Analysis}\label{subsec:time_complexity}
\begin{figure}[htbp]
\centering		
\includegraphics[scale=0.75]{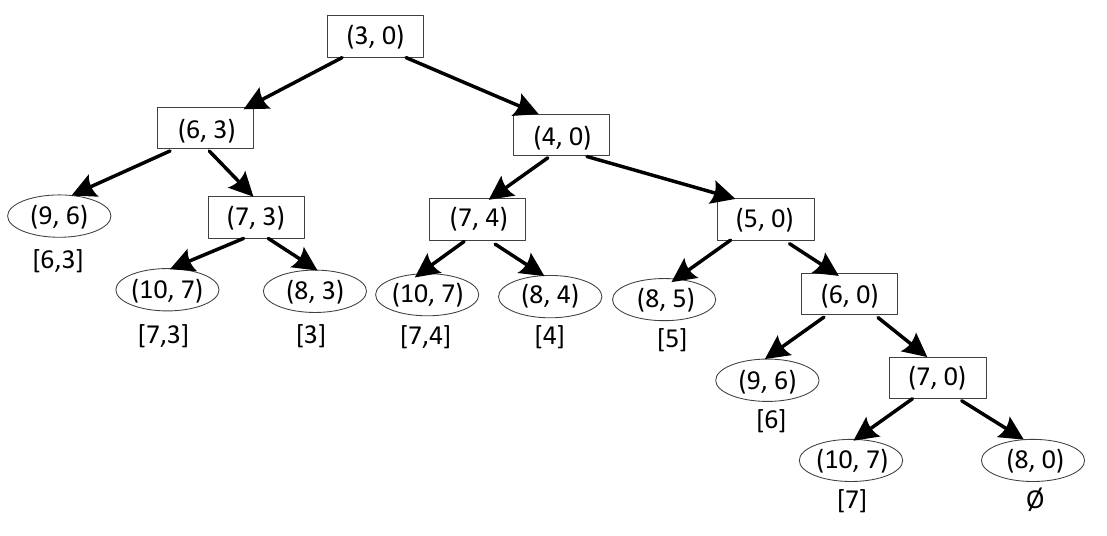}
\caption{\small Function calls of Algorithm~\ref{alg:error-distr} for a 32-bit
CSAA with $k=4$ and $l=8$.}
\label{fig:binary_tree}
\end{figure}

As shown in Algorithm~\ref{alg:error-distr}, the time complexity of the
proposed method is proportional to the number of calls of the function $ED$.
Since $ED$ is a recursive function and it calls itself twice in each function
call, the process of the function calls can be modeled as a binary tree. For
example, the binary tree representing the function calls for a 32-bit CSAA with
$k=4$ and $l=8$ is shown in Fig.~\ref{fig:binary_tree}. Each pair $(x, y)$ in
this figure represents one function call of the form $ED(i, j, ePar, pPar)$,
where $x$ is equal to $i$ and $y$ is equal to $j$. The internal nodes in this
binary tree are represented by square nodes, while the leaf nodes are
represented by round nodes. At the leaf, there is no further recursive function
call, since $i \ge m = n/k = 32/4= 8$. The vector under each leaf node
represents the error pattern produced at that leaf. A number $j$ in the vector
indicates that there is a 1 at the bit position $jk$ of the error pattern. For
example, the vector $[6,3]$ represents an error pattern with 1s at bit
positions $24$ and $12$. Note that the rightmost leaf corresponds to an
all-zero error pattern. Thus, the vector is $\emptyset$.
We can see that the number of error patterns is equal to the number of leaves
in the binary tree. Denote the number of error pattern as $N$.  Then the total
number of nodes in the binary tree is $2N-1$. Since each call of $ED$
corresponds to a node in the binary tree, the runtime of the proposed method is
proportional to $2N-1$, which implies that the time complexity is $O(N)$.  On
the other hand, any algorithm for obtaining the error distribution must have
runtime complexity at least $\Omega(N)$, since it must produce all the error
patterns and the associated probabilities. Therefore, the proposed algorithm
achieves the theoretical lower bound on the asymptotic runtime.

To further quantify the runtime, we analyze the number of error patterns for a
given block-based approximate adder. We show how we derive the number for a
block-based approximate adder using an example. Consider an adder with $n=32$,
$k=4$, and $l=4$. The number of blocks $m=n/k=8$ and $t=l/k=1$.  The following
is a list of all the non-zero error patterns, represented using the vector
representation shown in Fig.~\ref{fig:binary_tree}:
\begin{align*}
&[7],\\
&[6],\\
&[5], [7,5],\\
&[4], [6,4], [7,4],\\
&[3], [5,3], [7,5,3], [6,3], [7,3],\\
&[2], [4,2], [6,4,2], [7,4,2], [5,2], [7,5,2], [6,2], [7,2].
\end{align*}
Note that the error patterns in the same row above have their lowest 1 at
the same bit position. For example, in the $4$-th row, all the three error
patterns have their lowest 1 at the bit position $4k=16$.

If we denote the number of error patterns with lowest 1 at bit position $ik$
as $x_{m-i}$, we have $x_1=x_2=1$, $x_3=2$, $x_4=3$, $x_5=5$, and $x_6=8$ for
this example. As we showed in Sections~\ref{subsec:mul}, the lowest bit
position in an error pattern where a 1 can occur is $(t+1)k$. Therefore, the
maximal index of the sequence $x_i$ is $m-t-1=6$. Based on the numerical values
of $x_i$'s, we find that $x_i$'s satisfy the following pattern:
\begin{equation}
\label{eqn:x-i-1}
x_i = 1,
\end{equation}
for all $1 \leq i \leq 2$, and
\begin{equation}
\label{eqn:x-i-recursive}
x_i=x_{i-2}+x_{i-3}+\cdots+x_1+1,
\end{equation}
for all $2 < i \le 6$. Indeed, the above two equations make sense.  For any $1
\le i \le t+1=2$, if an error pattern has its lowest 1 at bit position
$(m-i)k$, then by Theorem~\ref{thm:error-distribution}, it cannot have any
1s on the left of that bit position. Therefore, the number of error patterns
with the lowest 1 at bit position $(m-i)k$ is just 1, and hence
Eq.~\eqref{eqn:x-i-1} holds. For any $2 < i \le 6$, the set of error patterns
with the lowest 1 at bit position $(m-i)k$ can be partitioned into $(i-1)$
subsets. The subset $j$ ($1 \le j \le i-2$) contains all the error patterns
with their second lowest 1 at bit position $(m-i+1+j)k$. The subset $(i-1)$
contains all the error patterns with no 1 on the left of bit position $(m-i)k$.
The number of error patterns in the subset $j$ ($1 \le j \le i-2$) is equal to
the number of error patterns with the lowest 1 at bit position $(m-i+1+j)k$.
According to the definition, this number is just $x_{i-j-1}$. The number of
error patterns in the subset $(i-1)$ is just 1. Therefore, we have
\begin{equation*}
x_i = \sum_{j=1}^{i-2} x_{i-j-1} + 1 = x_{i-2} + x_{i-3} + \cdots + x_1 + 1.
\end{equation*}
Thus, Eq.~\eqref{eqn:x-i-recursive} also holds.

Indeed, in the general case, no matter whether $l$ is a multiple of $k$ or not,
we have
\begin{equation}
\label{eqn:x-i-1-general}
x_i = 1,
\end{equation}
for all $1 \leq i \leq t+1$, and
\begin{equation}
\label{eqn:x-i-recursive-general}
x_i=x_{i-t-1}+x_{i-t-2}+\cdots+x_1+1,
\end{equation}
for all $t+1 < i \le m-t-1$. Note that the total number of non-zero error
patterns is given by $x_{m-t-1}+x_{m-t-2}+\ldots+x_1$. To count the total
number of error patterns, we should also include the zero error pattern.
Therefore, the total number of error patterns is
\begin{equation*}
N = x_{m-t-1}+x_{m-t-2}+\ldots+x_1 + 1.
\end{equation*}
If we extend the recursive definition of $x_i$ shown in
Eq.~\eqref{eqn:x-i-recursive-general} also to $i > m-t-1$, then the number of
error patterns is just $x_m$.

Note that by Eq.~\eqref{eqn:x-i-recursive-general}, we also have
\begin{equation}
\label{eqn:x-i-recursive-general-2}
x_i=x_{i-t-1}+x_{i-1},
\end{equation}
for all $i > t+1$.

In summary, the number of error patterns is given by $x_m$, where $x_i$ ($i
\ge 1$) is recursively defined by Eq.~\eqref{eqn:x-i-1-general}
and~\eqref{eqn:x-i-recursive-general-2}. When $t=0$, we have $x_m = 2^{m-1}$.
Therefore, the runtime of the proposed algorithm is $O(2^m)$.
When $t=1$, then the sequence $x_i$ is a Fibonacci sequence and $x_m$ is the
$m$-th value in the sequence, which is equal to $\frac{1}{\sqrt{5}} \left(
\left( \frac{1+\sqrt{5}}{2} \right)^m - \left( \frac{1-\sqrt{5}}{2} \right)^m
\right)$. Therefore, the runtime of the proposed algorithm is
$O\left(\left(\frac{\sqrt{5}+1}{2}\right)^m\right)$.

\section{Experimental Results}
\label{sec:experiment}

We implemented our methods for calculating error rate and error distribution
using C++. The error distributions of several block-based approximate adders
obtained by our method are shown in Section~\ref{subsec:ED}. Analysis was made
to explain the special pattern of the error distributions. We compared the
runtime and accuracy of the proposed method to other existing methods in
Sections~\ref{subsec:RT} and~\ref{subsec:ACC}, respectively.

\subsection{Error Distribution Study}\label{subsec:ED}
\begin{figure*}[!t]
\centering
\begin{tabular}{cc}
\subfloat[\label{subfig:err-dist-eta4}]{\includegraphics[scale = 1]{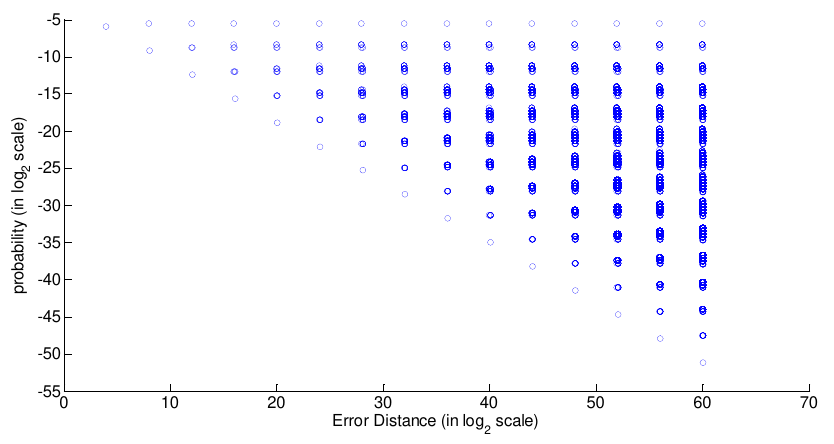} }&
\subfloat[\label{subfig:err-dist-eta2}]{\includegraphics[scale = 1]{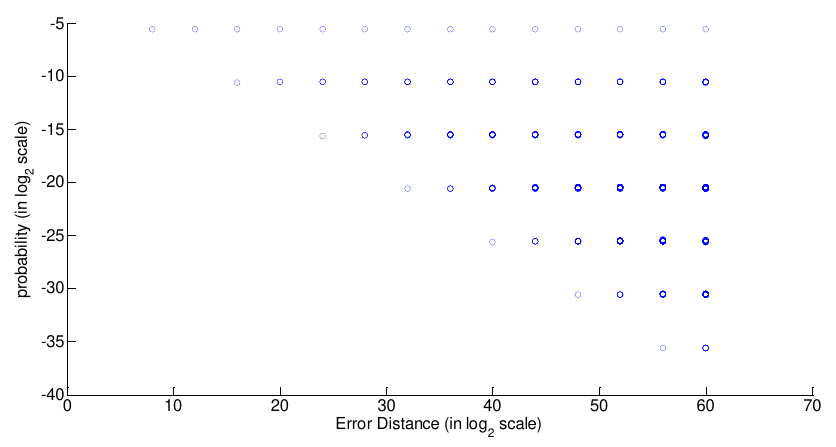}} \\
\subfloat[\label{subfig:err-dist-csa}]{\includegraphics[scale = 1]{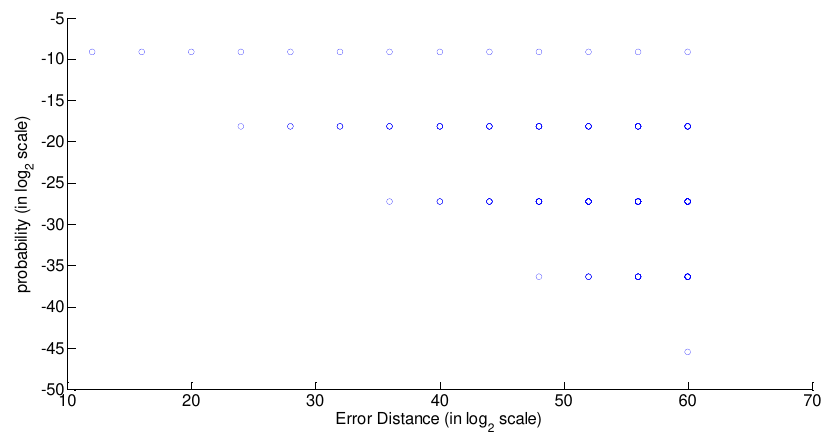} }&
\subfloat[\label{subfig:err-dist-k4l10}]{\includegraphics[scale = 1]{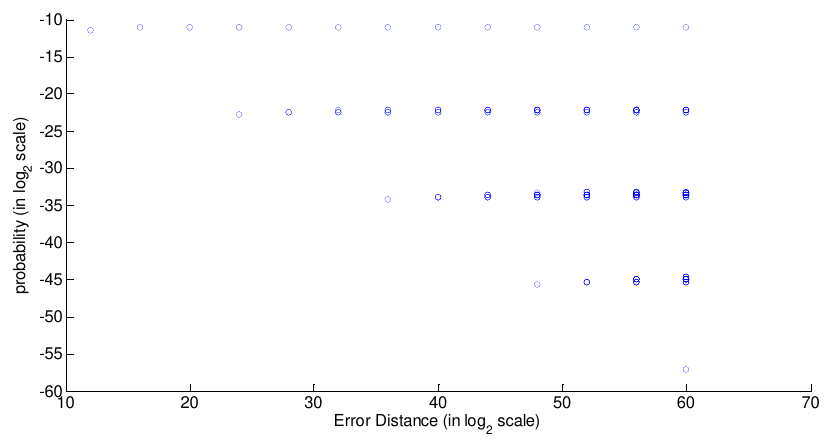}}
\end{tabular}
\caption{\small Scatter diagrams for error distribution produced by our method
for: (a) ETA-IV with $k=4$ and $l=2$; (b) ETA-II with $k=4$ and $l=4$; (c) CSAA
with $k=4$ and $l=8$; (d) a block-based approximate adder with $k=4$ and
$l=10$.}
\label{fig:err-dist}
\end{figure*}

Exact error distributions were generated for four block-based approximate
adders using our method, which are ETA-IV with $k=4$ and $l=2$, ETA-II with
$k=4$ and $l=4$, CSAA with $k=4$ and $l=8$, and a block-based approximate adder
with $k=4$ and $l=10$. All these adders are 64-bit in size. The scatter
diagrams of the error distributions of the four adders are shown in
Fig.~\ref{fig:err-dist}.

In Fig.~\ref{fig:err-dist}, each circle represents an error distance and its
associated probability. In each plot, the $x$-axis is the error distance in
the $\log_2$ scale and the $y$-axis is the probability also in the $\log_2$
scale. We can see that in each figure, the scatter points approximately form a
triangle.  Inside the triangle, points are located in a regular way.  However,
if we zoom in the figures, we will find that some circles overlap with others.
The circles with a thicker outline or the solid rounded rectangles in the
figures are actually resulted by the clustering of a number of circles. The
reason for the clustering along the $x$-axis is that the error distance of an
error pattern is dominated by the highest 1 in the error pattern. For example,
if the highest 1 of an error pattern appears at the bit position $36$, then the
1s on the right of bit position $36$ in the error pattern can only appear at
bit position lower than or equal to $32$. Thus, the error distance is close to
$2^{36}$ (indeed, the error distance is larger than or equal to $2^{36}$),
which are approximately 36 in the $\log_2$ scale. Thus, all error patterns
which have their leading 1s at the same location are located very closely in
the $x$ dimension. Since the leading 1 in an error pattern can only occur at
bit positions $ik$, where $k=4$ and $i$ is an integer, the clusters are located
near $x = 4i$ in the $\log_2$ scale, as shown in the figures.

Moreover, we can also observe from the figures that for points with close error
distances, they cluster into a number of groups along the $y$-axis. The reason
behind this is that among the error patterns with the leading 1 at the same
location, the patterns with the same number of 1s have similar probabilities.
We use the $64$-bit CSAA with $k=4$ and $l=8$ to illustrate this. Consider all
error patterns with their error distances around $36$ in the $\log_2$ scale.
Their leading 1s are all at bit position $36$. Since $t = l/k = 2$, two
adjacent 1s in any error pattern must be at least $12$ bits (or $3$ blocks)
away and the lowest bit position that can have a 1 is $12$. Thus, there are at
most three 1s in these error patterns. We divide them into three sets based
on the number of 1s in the error pattern.  The first set consists of the
error patterns with a single 1. Using the vector representation shown in
Fig.~\ref{fig:binary_tree}, the set includes only one error pattern of the
form $[9]$. By Theorem~\ref{thm:error-distribution}, its probability is
$d_6\cdot e_{9,0}$.  The second set is composed of error patterns with two
1s. They are $[9,6]$, $[9,5]$, $[9,4]$, and $[9,3]$.  Their probabilities are
$d_6\cdot(e_{9,6}e_{6,0})$, $d_6\cdot(e_{9,5}e_{5,0})$,
$d_6\cdot(e_{9,4}e_{4,0})$, and $d_6\cdot(e_{9,3}e_{3,0})$, respectively. The
third set consists of error patterns with three 1s. The set only has one
element, which is $[9,6,3]$. Its probability is
$d_6\cdot(e_{9,6}e_{6,3}e_{3,0})$. It can be seen that the probability of an
error pattern with $q$ 1s has $q$ $e_{i,j}$ terms in its product expression.

\begin{figure}[!t]
\centering
\includegraphics[scale=0.8]{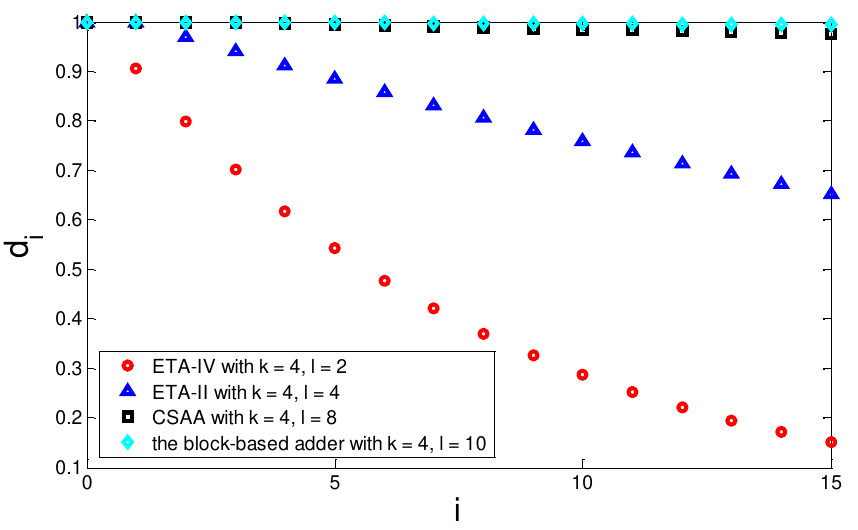}		
\caption{\small $d$ values for ETA-IV with $k=4$ and $l=2$, ETA-II with $k=4$
and $l=4$, CSAA with $k=4$ and $l=8$, and a block-based approximate adder with
$k=4$ and $l=10$.}
\label{fig:d_value}
\end{figure}

Now consider $e_{i,j}$. Since for the example approximate adder we consider,
its $l$ is a multiple of $k$,  the value $e_{i,j}$ is given by
Eq.~\eqref{eqn:e-i-j}, i.e., $e_{i,j} = P(P_{i-1}) \cdots P(P_{i-t})
P(G_{i-t-1}) d_{i-t-j-1}$. Under the assumption that the inputs are uniformly
distributed, the value $P(P_{i-1}) \cdots P(P_{i-t}) P(G_{i-t-1})$ is a
constant for a fixed $t$. We denote this constant as $c$. Then, we can
represent the value of $e_{i, j}$ as $c \cdot d_{i-t-j-1}$. Therefore, the
value of $e_{i,j}$ is proportional to $d_{i-t-j-1}$. The sequence of black dots
in Fig.~\ref{fig:d_value} shows the values of $d_i$ for $0 \le i \le m-1$ for
the CSAA with $k=4$ and $l=8$.  From the figure, we can see that all $d_i$'s
are very close. As a result, all $e_{i,j}$'s for $0 \le i,j \le m-1$ and $i-j >
t$ are also very close.  Consequently, we can see that the error patterns  in
the same set have very close probabilities, since their probability
expressions contain the same number of $e_{i,j}$ terms. Therefore, the error
patterns in the same set also cluster in the $y$ dimension. Since there are
three sets, the number of clusters is three, as shown in
Fig.~\ref{subfig:err-dist-csa}. Furthermore, since the number of $e_{i,j}$
terms in the probability expression of an error pattern is equal to the number
of 1s in the error pattern, the probability value of an error pattern in the
$\log_2$ scale is proportional to the number of 1s in the error pattern. This
explains why the distances along the $y$ dimension between any two adjacent
clusters are the same.  Similar analysis can be made for points with other
error distances.

However, we found that the clustering of the points along the $y$-axis is
not so strong for ETA-IV, as shown in Fig.~\ref{subfig:err-dist-eta4}.
The reason lies in that $d_i$'s for $i = 0, 1, \cdots, m-1$ of this adder are
quite different from each other, as shown by the sequence of red dots in
Fig.~\ref{fig:d_value}, making the values of $e_{i, j}$ also vary greatly for
different $(i-j)$ ($i-j > t$).  Hence, for error patterns with close error
distances, although some of them have the same number of 1s, their
probabilities are not very close.


Based on the above analysis, we can see that the number of clusters over the
error patterns with close error distance (i.e., error patterns with the leading
1 at the same bit position) is equal to the maximal number of 1s in such error
patterns. Suppose the leading 1 is at bit position $ik$. By
Theorem~\ref{thm:error-distribution}, the maximal number of 1s in such error
patterns is $\lfloor i/(t+1) \rfloor$. Therefore, the number of clusters on the
error patterns with error distance close to $ik$ (in the $\log_2$ scale) is
equal to $\lfloor i/(t+1) \rfloor$. This explains the trend of how the number
of clusters increases with the error distance $ik$ as shown in the figures. For
example, for the ETA-II, as shown in Fig.~\ref{subfig:err-dist-eta2}, starting
from the first error distance $4\cdot 2$, two adjacent error distances $4i$ and
$4(i+1)$ have the same number of clusters. Then, the next two adjacent error
distances $4(i+2)$ and $4(i+3)$ have one more cluster than the previous two.
This pattern continues until the last error distance. This is because for
ETA-II, $t=1$ and hence the number of clusters for an error distance $ik$ is
equal to $\lfloor i/2 \rfloor$. From the figures, we can also see that the
minimum non-zero error distances for the four approximate adders are different,
while the maximal error distances are close. The minimal non-zero error
distances for ETA-IV, ETA-II, CSAA, and the block-based approximate adder with
$k=4$ and $l=10$ are $4$, $8$, $12$, and $12$ in the $\log_2$ scale,
respectively, while the maximal error distances are all close to $60$. This is
reasonable, since by Theorem~\ref{thm:error-distribution}, for a non-zero error
pattern, the lowest and the highest bit positions where a 1 can occur are
$(t+1)k$ and $(m-1)k$, respectively.


\begin{figure*}[htbp]
\centering
\begin{tabular}{cccc}			
\subfloat[\label{subfig:err-bar-eta4}]{\includegraphics[width=0.23\textwidth]{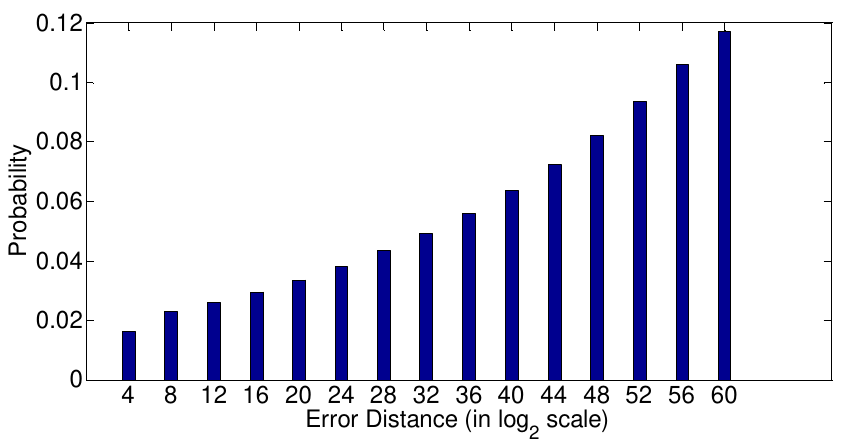} }&
\subfloat[\label{subfig:err-bar-eta2}]{\includegraphics[width=0.23\textwidth]{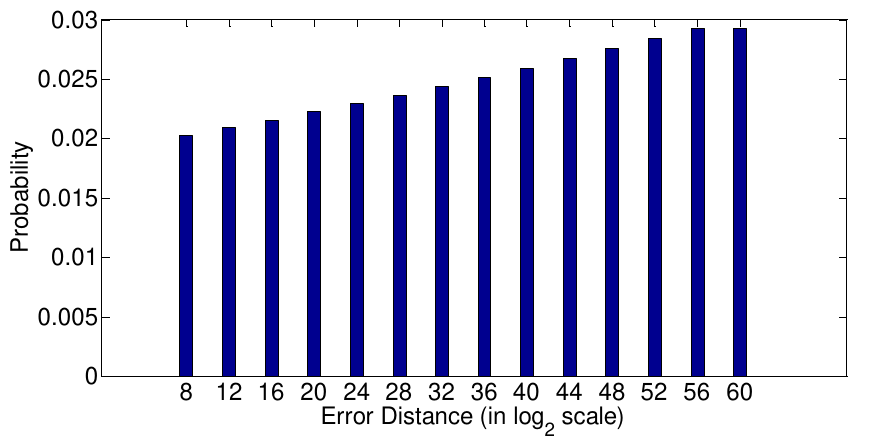}} &
\subfloat[\label{subfig:err-bar-csa}]{\includegraphics[width=0.23\textwidth]{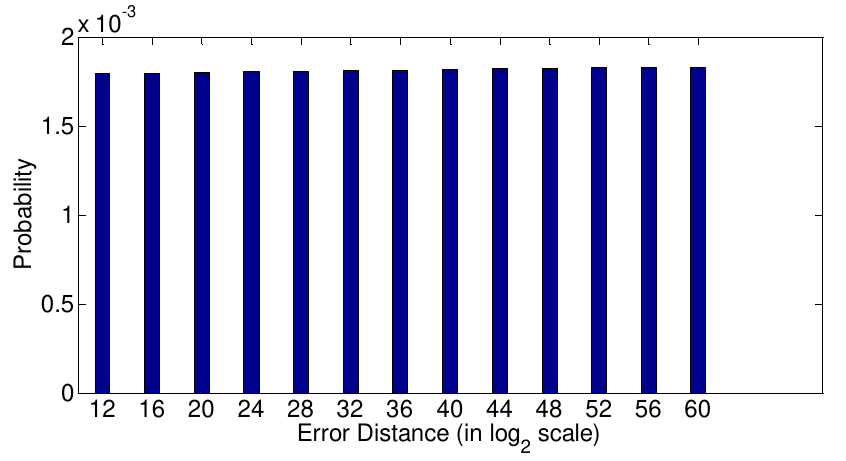} }&
\subfloat[\label{subfig:err-bar-k4l10}]{\includegraphics[width=0.23\textwidth]{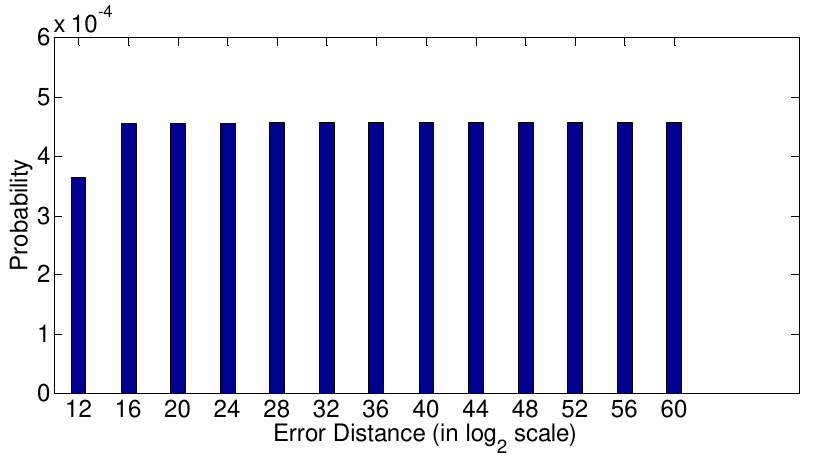}}
\end{tabular}
\caption{\small Distributions on the sum of probabilities of all the error
patterns with close error distance for: (a) ETA-IV with $k=4$ and $l=2$; (b)
ETA-II with $k=4$ and $l=4$; (c) CSAA with $k=4$ and $l=8$; (d) a block-based
approximate adder with $k=4$ and $l=10$.} \label{fig:err-dist-bar}
\end{figure*}

Given that the error patterns are clustered around error distance $ik$ in the
$\log_2$ scale, we are also interested in studying how the sum of the
probabilities of all the error patterns with distance around $ik$ changes with
the value $ik$. For the four approximate adders above, we obtained the sums and
plotted the bar graphs shown in Fig.~\ref{fig:err-dist-bar}.
It can be easily seen that the height of each bar located at error distance
$ik$ (in the $\log_2$ scale) also represents the probability of an error
pattern with the leading 1 at the bit position $ik$, or the probability of an
error pattern with error distance in the range $[2^{ik}, 2^{(i+1)k})$. We
denote such a probability as $P(LO=ik)$. From the figures, we can see that as
$i$ increases, the probability $P(LO=ik)$ also increases or keeps unchanged.
However, for different adders, the trends are different. To understand the
reason, we will analyze the probability $P(LO=ik)$.  Indeed, we have $$P(LO=ik)
= P(e[>ik] = 0|e[ik]=1) P(e[ik]=1),$$ where $P(e[>ik] = 0|e[ik]=1)$ is the
probability that there is no 1 on the left of the bit position $ik$ under the
condition that there is a 1 at bit position $ik$, and $P(e[ik]=1)$ is the
probability that there is a 1 at the bit position $ik$. From
Section~\ref{sec:error-distr}, we know that $P(e[>ik] = 0|e[ik]=1) =
d_{m-i-1}$. Now we only need to obtain $P(e[ik]=1)$.

\begin{figure*}[t]
\centering
\begin{tabular}{cccc}			
\subfloat[\label{subfig:rt-l2}]{\includegraphics[width=0.23\textwidth]{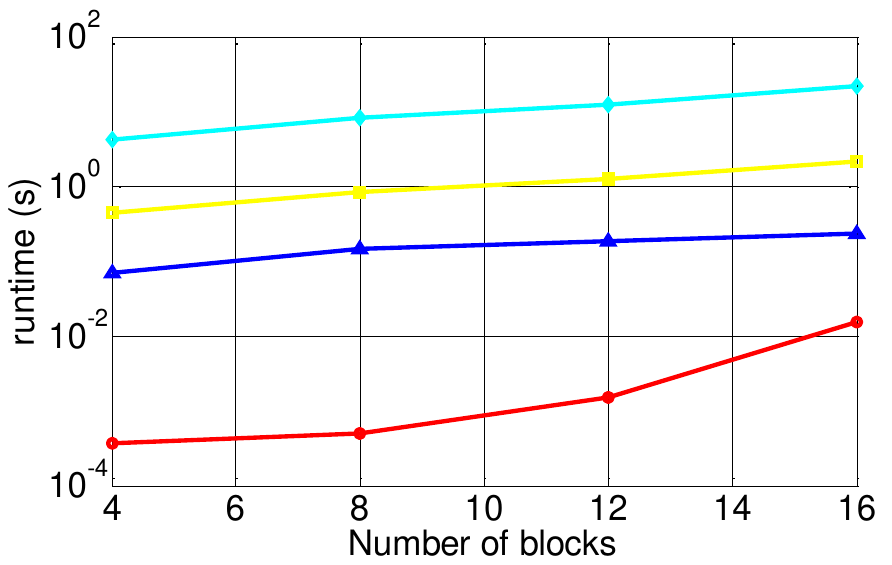} } &
\subfloat[\label{subfig:rt-l4}]{\includegraphics[width=0.23\textwidth]{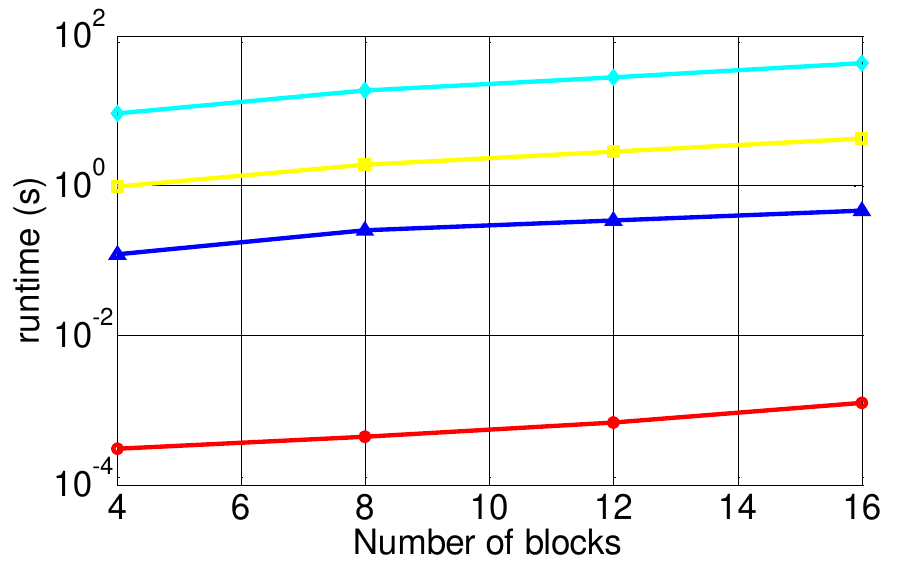}} &
\subfloat[\label{subfig:rt-l8}]{\includegraphics[width=0.23\textwidth]{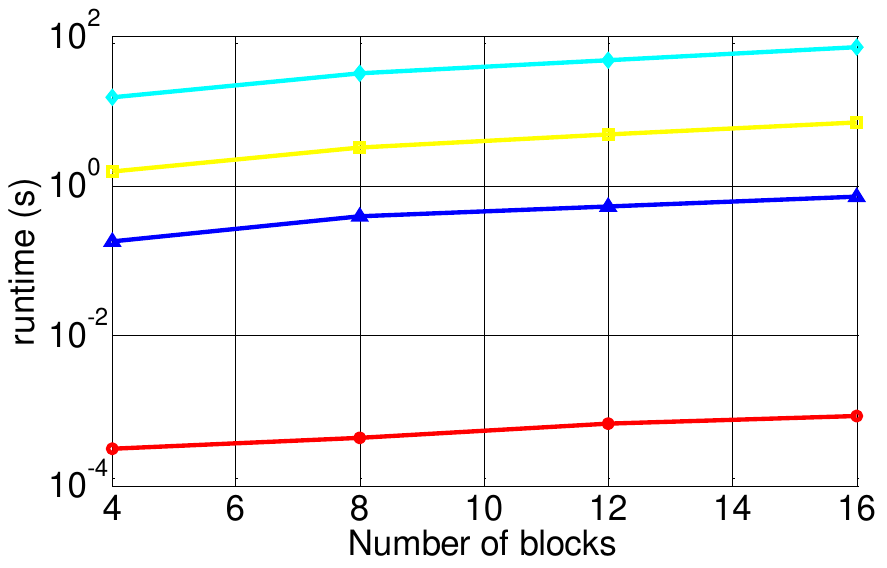}} &
\subfloat[\label{subfig:rt-l10}]{\includegraphics[width=0.23\textwidth]{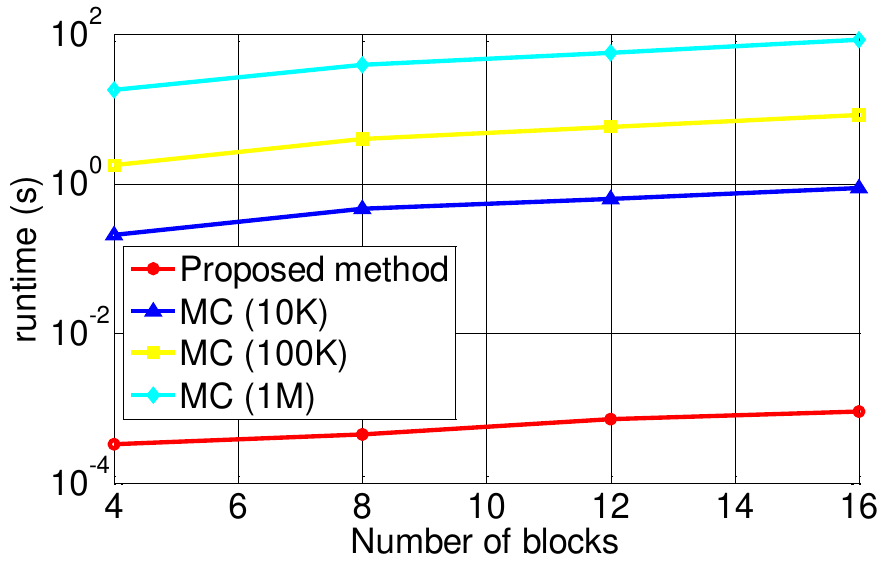}}
\end{tabular}
\caption{\small Runtime comparison between the proposed method to obtain the
error distribution and the Monte-Carlo (MC) sampling method for: (a) ETA-IV
with $k = 4$ and $l = 2$; (b) ETA-II with $k = 4$ and $l = 4$; (c) CSAA with $k
= 4$ and $l = 8$; (d) a block-based approximate adder with $k = 4$ and $l =
10$.}
\label{fig:runtime}
\end{figure*}

If $l$ is a multiple of $k$, then by Theorem~\ref{thm:err-cond}, $P(e[ik]=1) =
P(P_{i-1}) \cdots P(P_{i-t}) P(G_{i-t-1})$. Under the assumption that the
inputs are uniformly distributed, the product is a constant  for a fixed $t$
and hence, $P(LO=ik)$ is proportional to $d_{m-i-1}$. This can be verified by
Fig.~\ref{subfig:err-bar-eta2} and~\ref{subfig:err-bar-csa}, which show the bar
graphs for the ETA-II and CSAA, respectively. Indeed, the trend of how the
height of the bar changes with error distance $ik$ follows the same trend of
how the sequence $d_{m-t-1}, d_{m-t-2}, \ldots, d_0$ changes, which can be
observed from Fig.~\ref{fig:d_value}, in which the sequence of blue dots
corresponds to the sequence $d_0, d_1, \ldots, d_{m-1}$ for the ETA-II and the
sequence of black dots corresponds to the sequence for CSAA. As shown in
Fig.~\ref{fig:d_value}, for CSAA, the sequence $d_0, \ldots, d_{m-1}$ decreases
very slowly. This explains why the bars in the graph for CSAA increases very
slowly.

If $l$ is not a multiple of $k$, then $P(e[ik]=1)$ can be obtained by
Theorem~\ref{thm:err-cond-k-neq}.  However, in this case, $P(e[ik]=1)$ for
$i=t+1$ differs from $P(e[ik]=1)$'s for $i > t+1$. Indeed, we have
\begin{equation*}
P(e[ik]=1)=
\begin{cases}
A(i), & i = t+1 \\
A(i)+B(i), & i > t+1
\end{cases},
\end{equation*}
where $A(i) = P(P_{i-1}) \cdots P(P_{i-t}) P(PL_{i-t-1}) P(GR_{i-t-1})$ and
$B(i)= P(P_{i-1}) \cdots P(P_{i-t-1}) P(GL_{i-t-2})$. Under the assumption that
the inputs are uniformly distributed, both $A(i)$ and $B(i)$ are positive
constants independent of $i$. This explains the special pattern shown in
Fig.~\ref{subfig:err-bar-k4l10} for the block-based approximate adder with
$k=4$ and $l=10$. From the figure, we can see the heights of the second bar up
to the last bar are almost the same. This is because $P(e[ik]=1)$ are the same
for all $i>t+1$ and the values $d_0, d_1, \ldots, d_{m-1}$ are very close as
shown by the sequence of cyan dots in Fig.~\ref{fig:d_value}. However, we can
see the height of the first bar is obviously shorter than the other bars.  This
is because $P(e[(t+1)k]=1)$ is smaller than any other $P(e[ik]=1)$ ($i > t+1$)
by a positive constant $B(i)$. The above reasoning can be also applied to
explain the pattern shown in Fig.~\ref{subfig:err-bar-eta4} for the ETA-IV.
Specifically, the trend from the second bar to the last bar follows the trend
of the sequence $d_{m-3}, d_{m-4}, \ldots, d_{0}$, which is shown by the
sequence of red dots in Fig.~\ref{fig:d_value}. The first bar is short than the
second bar because $P(e[(t+1)k]=1)$ is smaller than $P(e[(t+2)k]=1)$ and
$d_{m-2} < d_{m-3}$.

\subsection{Runtime Study}\label{subsec:RT}

In this section, we compared the runtime of our method to obtain error
distribution with the Monte Carlo sampling method, which randomly chooses a
subset of input combinations for simulation, and the exhaustive method, which
enumerates all the input combinations.  All the methods were implemented in
{\tt C++}. All the experiments were conducted on a virtual machine running
Linux operating system with 1GB memory.  The host machine is a 3.1 GHz desktop.
We also compared the asymptotic runtime of our method with the exact method
proposed in~\cite{Mazahir16}.


Fig.~\ref{fig:runtime} shows the runtime of the proposed method to obtain error
distribution and the Monte-Carlo methods with 10K, 100K, and 1M samples on four
approximate adders: ETA-IV with $k=4$ and $l=2$, ETA-II with $k=l=4$, CSAA with
$k=4$ and $l=8$, and a block-based approximate adder with $k=4$ and $l=10$. For
each type of adder, we did experiments on four different block numbers $m= 4,
8, 12, 16$. Thus, the operand sizes $n$ vary from $16$ to $64$.

From the four plots in Fig.~\ref{fig:runtime}, we can see that the proposed
method consumes much less time than the Monte-Carlo method. For all the
experiments, the proposed method takes less than 0.02 second. For three
$64$-bit approximate adders with $l = 4$, $l = 8$, and $l = 10$, the proposed
method needs less than 0.002 seconds. In contrast, for the Monte-Carlo method
with 10K samples, the runtime is $0.05 \sim 0.9$ seconds. As the sample size
increases, the runtime of the Monte Carlo method also increases linearly. When
the sample size reaches 1M, the runtime of the Monte-Carlo method is $5 \sim
90$ seconds.

Comparing the four plots in Fig.~\ref{fig:runtime}, we can also see that the
runtime of the Monte-Carlo method increases with the carry generator size $l$.
This is reasonable since a larger carry generator size $l$ implies longer
simulation time. In contrast, the runtime of the proposed method decreases with
$l$. Especially, our method for ETA-IV with $l = 2$ takes much more time than
the other three adders when the number of blocks are 12 and 16.  This is also
reasonable. As we discussed in Section~\ref{subsec:time_complexity}, the
runtime of the proposed method is proportional to $N$, where $N$ is the total
number of error patterns of a block-based approximate adder. For a fixed $n$
and $k$, as $l$ decreases, the value $t = \lfloor l/k \rfloor$ decreases.
Consequently, the number of error patterns increases.  According to our
experimental results, a $64$-bit ETA-IV with $k=4$ and $l=2$ has 32768 error
patterns, while a $64$-bit ETA-II with $k=4$ and $l=4$ has 987 error patterns.
The number of error patterns for a $64$-bit CSAA with $k=4$ and $l=8$ and a
$64$-bit block-based approximate adder with $k=4$ and $l=10$ are both 189.
Therefore, the $64$-bit ETA-IV consumes much more time than the other three
adders.

\begin{table}[!t]
\caption{\small Runtime of the exhaustive method in second.}
\begin{center}
\label{tab:rt-enumeration}
\begin{tabular}{c|ccc}
\hline
 & $n = 8$ & $n = 12$  & $n = 16$\\
\hline
ETA-IV, $k = 4, l = 2$ & 0.09 & 30 & 7949\\
\hline
ETA-II, $k = 4, l = 4$ & 0.148 & 54 & 18877\\
\hline
CSAA, $k = 4, l = 8$ & 0.209 & 83 & 31340\\
\hline
Block-based, $k = 4, l = 10$ & 0.148 & 97 & 36668\\
\hline
\end{tabular}
\end{center}
\end{table}

To compare our method with the exhaustive method, we obtained the runtime of
the exhaustive method for the four types of approximate adders with size
$n=8,12,16$. The results are shown in Table~\ref{tab:rt-enumeration}.  We can
see that when $n = 16$, the runtime for the exhaustive method has already
reached 7000 seconds, which is much larger than our method. Indeed, the
exhaustive method has a time complexity of $O(4^n)$, where $n$ is the size of
an adder. As a result, its runtime explodes very quickly.

\begin{figure}[!t]
\centering
\includegraphics[scale=1]{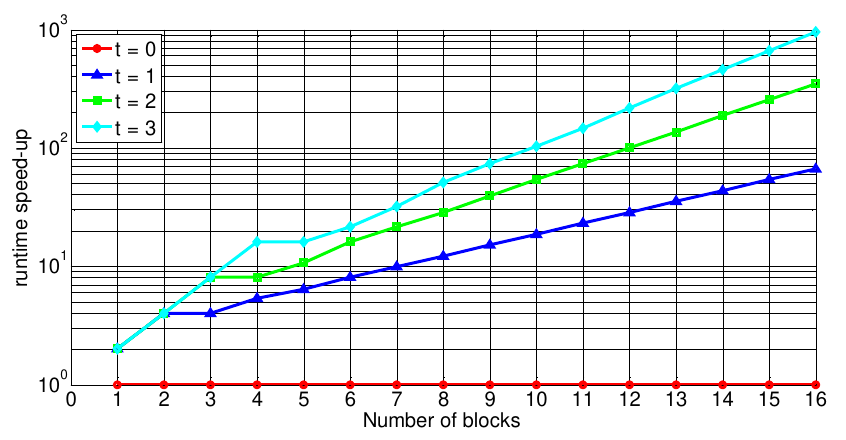}		
\caption{\small Asymptotic runtime speed-up of the proposed method over the
method in~\cite{Mazahir16}.}
\label{fig:runtime_complexity}
\end{figure}

Finally, we also compared the asymptotic runtime of our method with the method
proposed in~\cite{Mazahir16}. The asymptotic runtime of the method
in~\cite{Mazahir16} is $O(2^m)$ and that of our method is $O(N(m,t))$, where
$N(m,t)$ is the total number of error patterns of a block-based approximate
adder, determined by the number of blocks $m$ and the value $t = \lfloor l/k
\rfloor$. The value $N(m,t)$ can be obtained by the method shown in
Section~\ref{subsec:time_complexity}.  By ignoring the constant, we used the
ratio $2^m/N(m,t)$ to measure the asymptotic runtime speed-up of our method
over the method in~\cite{Mazahir16}. In Fig.~\ref{fig:runtime_complexity}, for
each $t=0,1,2,3$, we show the asymptotic runtime speed-up versus $m$ for $m$
ranging from $1$ to $16$. We can see that for block-based approximate adders
with $t=0$, such as ETA-IV, both methods have the same asymptotic runtime.
However, for adders with $t > 0$, such as ETA-II and CSAA, our method is much
more efficient than the method in~\cite{Mazahir16}. On the one hand, for any
fixed $t > 0$, the speed-up grows exponentially with $m$. On the other hand,
for any fixed $m$, the speed-up also increases with $t$, because the number of
error patterns drops as $t$ increases. From our experimental results, when
$m=16$, the speed-ups for $t = 1$, $2$, and $3$ are approximately 66, 347, and
950, respectively, demonstrating that our method is much more efficient than
the method~\cite{Mazahir16} in providing the exact error distributions for
block-based approximate adders.

\subsection{Accuracy Study}\label{subsec:ACC}
\begin{figure*}[!t]
\centering
\begin{tabular}{cccc}			
\subfloat[\label{subfig:range_k4l2}]{\includegraphics[width=0.23\textwidth]{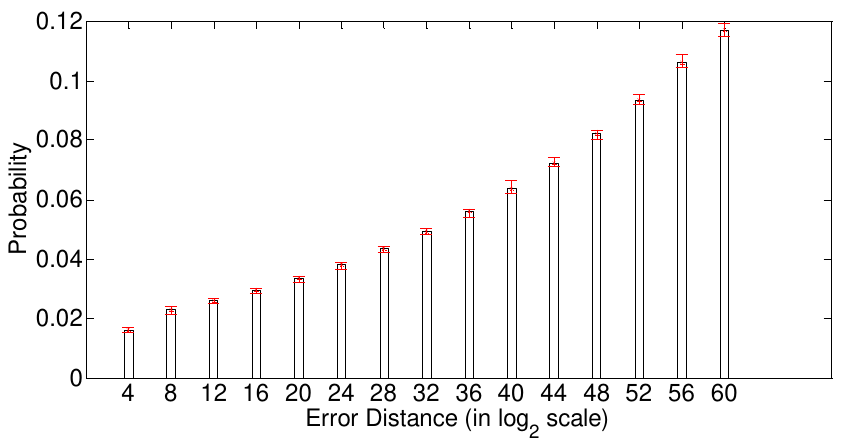} }&
\subfloat[\label{subfig:range_k4l4}]{\includegraphics[width=0.23\textwidth]{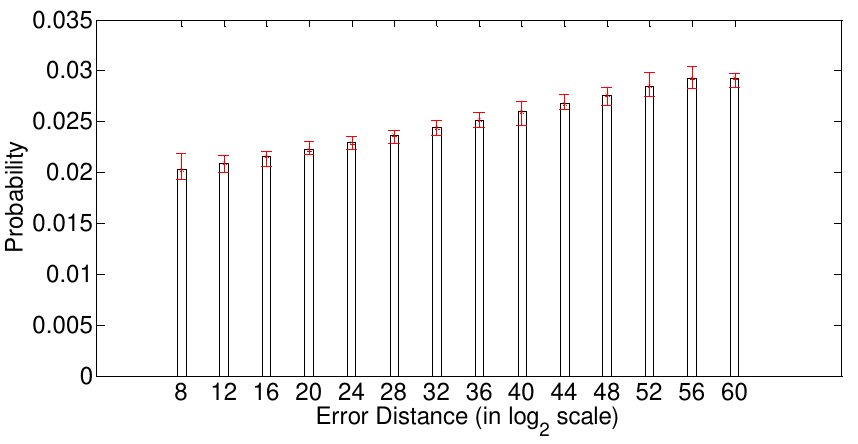}} &
\subfloat[\label{subfig:range_k4l8}]{\includegraphics[width=0.23\textwidth]{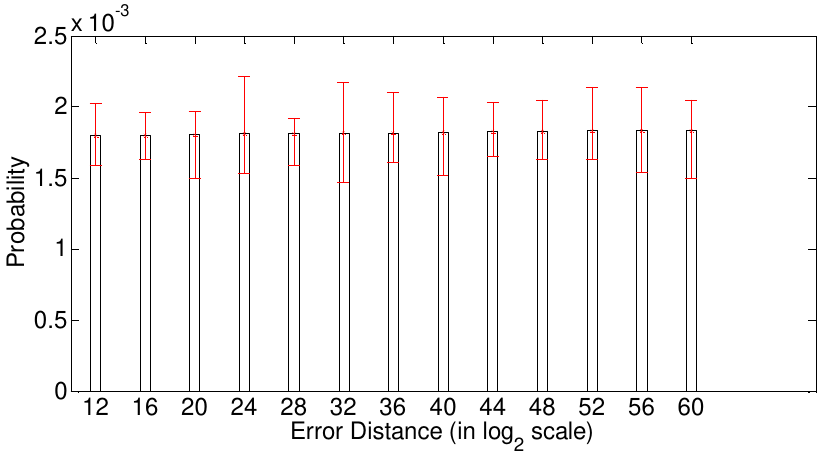} }&
\subfloat[\label{subfig:range_k4l10}]{\includegraphics[width=0.23\textwidth]{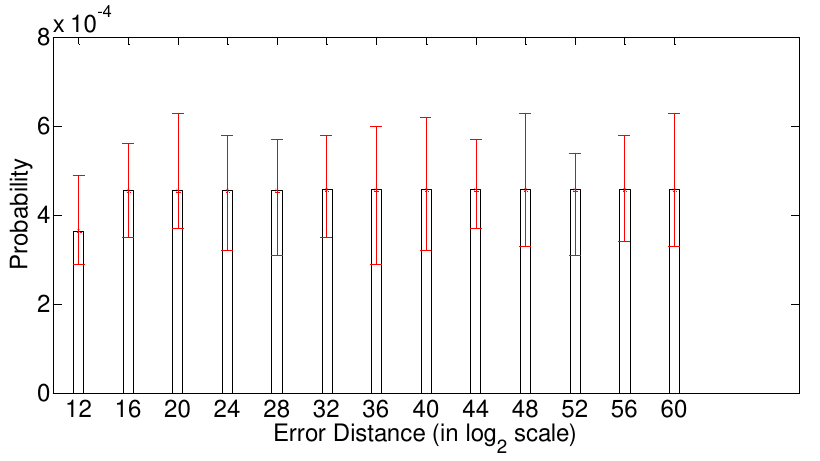}}
\end{tabular}
\caption{\small Comparison between our method and the Monte Carlo sampling
method with 100K samples in producing error distribution for: (a) ETA-IV with
$k=4$ and $l=2$; (b) ETA-II with $k=4$ and $l=4$; (c) CSAA with $k=4$ and
$l=8$; (d) a block-based approximate adder with $k=4$ and $l=10$.}
\label{fig:dist_comp}
\end{figure*}

Since the error distribution produced by our method is exact, we can further
obtain the exact error statistics, such as error rate (ER), mean error distance
(MED), and mean square error (MSE). In this section, we demonstrate the
advantage of our method in accuracy by comparing it with the method
in~\cite{li2014error} and the Monte-Carlo sampling method.

To show the advantage in accuracy, we used the results of the proposed method
as the reference and computed the relative errors of ERs and MSEs obtained by
the method in~\cite{li2014error} and the Monte Carlo sampling methods with
sample sizes as 10K, 100K, and 1M. Note that since the work~\cite{li2014error}
has given an exact analytical expression for the mean error distance (MED), we
did not consider the relative errors of MED in our experiment. The relative
errors of ERs and MSEs in percentage are shown in
Table~\ref{tab:relative-error}. The experiments were performed on eight
different $32$-bit block-based approximate adders.  The types of the adders and
their $k$ and $l$ values are listed in the first column of the table.  For each
sample size $M$ in the Monte Carlo simulation, we ran the entire $M$-sample
simulation 100 times and obtained the relative error for each simulation run.
The final relative error listed in the table is the average relative
error over $100$ runs. The results in the table show that the relative
errors on ERs and the MSEs computed by the method in~\cite{li2014error} can be
up to 4.7\% and 8.6\%, respectively. For the Monte-Carlo method with sample size
equal to 10K, the relative errors on ERs and MSEs can be up to 15.0\% and
32.8\%, respectively. When the sample size increases, the relative errors
decrease. However, simulations with a larger sample size takes longer time.

\begin{table}[!htbp]
\caption{\small Relative errors in percentage of the method proposed
in~\cite{li2014error} and the Monte Carlo method.}
\begin{center}
\label{tab:relative-error}
\begin{tabular}{c|c|c|ccc}
\hline
\multicolumn{2}{c|}{} & \multirow{2}{*}{\cite{li2014error}}  & \multicolumn{3}{c}{Monte Carlo simulation}\\
\cline{4-6}
\multicolumn{2}{c|}{}      &  & 10K & 100K & 1M \\
\hline
\multirow{2}{*}{\tabincell{c}{ETA-IV\\ $k=2, l=1$}} & ER & 0.812 & 0.188 & 0.064 & 0.055 \\
                                                    & MSE & 8.573 & 1.705 & 0.521 & 0.161  \\
\hline
\multirow{2}{*}{\tabincell{c}{ETA-II\\ $k=2, l=2$}} & ER & 4.658 & 0.436 & 0.133 & 0.041  \\
                                                    & MSE & 1.541 & 2.439 & 0.669 & 0.222  \\
\hline
\multirow{2}{*}{\tabincell{c}{CSAA\\ $k=2, l=4$}}   & ER & 3.713 & 1.374 & 0.367 & 0.144  \\
                                                    & MSE & 0.098 & 4.531 & 1.652 & 0.502  \\
\hline
\multirow{2}{*}{$k=2, l=5$}                         & ER & 0.142 & 1.843 & 0.582 & 0.211  \\
                                                    & MSE & 0.037 & 6.629 & 2.049 & 0.813  \\
                                                    \hline
\multirow{2}{*}{\tabincell{c}{ETA-IV\\ $k=4, l=2$}} & ER & 0.381 & 0.676 & 0.189 & 0.064 \\
                                                    & MSE & 1.254 & 2.236 & 0.706 & 0.231 \\
\hline
\multirow{2}{*}{\tabincell{c}{ETA-II\\ $k=4, l=4$}} & ER & 2.331 & 1.687 & 0.608 & 0.195  \\
                                                    & MSE & 0.025 & 4.771 & 1.355 & 0.472  \\
\hline
\multirow{2}{*}{\tabincell{c}{CSAA\\ $k=4, l=8$}} & ER & 0.256 & 8.544 & 2.719 & 0.831  \\
                                                    & MSE & 0 & 17.284 & 5.130 & 1.669  \\
\hline
\multirow{2}{*}{$k=4, l=10$}                        & ER & 4.093 & 14.991 & 4.634 & 1.619  \\
                                                    & MSE & 0 & 32.770 & 11.240 & 3.922  \\
\hline
\end{tabular}
\end{center}
\end{table}

It should be noted that the method in~\cite{li2014error} can only estimate the
values of ER and MSE. It cannot provide error distributions. However, the
Monte-Carlo method could generate an error distribution through random
simulation, although the result is not exact.  Next, we compared the error
distributions given by the Monte-Carlo method to the exact error distributions
produced by the proposed method. We applied both methods to generate the type
of distribution as shown in Fig.~\ref{fig:err-dist-bar}. The results on
$64$-bit ETA-IV with $k = 4$ and $l = 2$, ETA-II with $k = 4$ and $l = 4$, CSAA
with $k = 4$ and $l = 8$, and a block-based approximate adder with $k = 4$ and
$l = 10$ are shown in Fig.~\ref{fig:dist_comp}. The white bars show the
accurate error distribution produced by our method, while the red lines on each
bar indicate the range of probabilities in 20 Monte Carlo simulation runs, each
with 100K samples. From the figure, we can see that when $l$ is small, the
Monte Carlo simulation can produce a result close to the accurate distribution,
but when $l$ increases, the accuracy of the Monte Carlo simulation degrades.
The reason is that when $l$ increases, the error probability decreases. For
example, as shown in Fig.~\ref{fig:dist_comp}, the error probability of ETA-IV
is on the scale of 0.01, while the error probability of the block-based
approximate adder with $k = 4$ and $l = 10$ is on the scale of 0.0001. As the
probability becomes smaller, the relative variation of the Monte Carlo result
becomes larger. This clearly indicates the accuracy of the Monte Carlo method
is reduced when the error probability is small, which is not uncommon for
approximate adders. In contrast, our method is guaranteed to obtain the exact
distribution.

\section{Conclusion}
\label{sec:conclusion}

In this paper, we proposed an accurate and efficient method to obtain the error
rate and error distribution of block-based approximate adders. The method to
obtain error distribution indeed achieves the theoretical lower bound on the
asymptotic runtime. Once the distribution is known, some other error metrics of
interest, such as mean error distance and mean square error, can be easily
obtained. Experimental results demonstrated that the proposed methods are
accurate and efficient.  Compared to the Monte Carlo sampling method, our
method is much better, especially when the error probability is small. 

In this work, we considered block-based approximate adders with the speculated
carry-in to the carry generator as $0$.  It is also possible to use the input
bit at one bit position lower as the speculated
carry-in~\cite{huang2012methodology}. In our future work, we will develop
techniques to analyze the error statistics for this type of approximate adders.

\section*{Acknowledgments}

This work is supported by National Natural Science Foundation of China (NSFC)
under Grant No. 61574089 and 61472243.

\bibliographystyle{IEEEtran}
\bibliography{references}

\end{document}